\documentclass[%
reprint,
 prd,
superscriptaddress,
nofootinbib,
amsmath,amssymb,  amsthm, amsfonts, latexsym,
aps
]{revtex4-2}
\usepackage[utf8]{inputenc}

\usepackage{amsmath, amsthm, amsfonts,  amssymb, latexsym}
\usepackage{graphicx}
\usepackage{dcolumn}
\usepackage{bm}
\usepackage{hyperref}
\usepackage[mathlines]{lineno}

\usepackage{enumerate}
\usepackage{mathrsfs}
\usepackage{graphicx}

\usepackage{subcaption}
\usepackage{xfrac}
\usepackage{lmodern}

\newcommand{\ri}{\mathrm{in}}
\newcommand{\ru}{\mathrm{up}}
\newcommand{\vol}{\mathrm{vol}}
\newcommand{\ud}{\mathrm{d}}
\DeclareMathOperator*{\SumInt}{%
	\mathchoice%
	{\ooalign{$\displaystyle\sum$\cr\hidewidth$\displaystyle\int$\hidewidth\cr}}
	{\ooalign{\raisebox{.14\height}{\scalebox{.7}{$\textstyle\sum$}}\cr\hidewidth$\textstyle\int$\hidewidth\cr}}
	{\ooalign{\raisebox{.2\height}{\scalebox{.6}{$\scriptstyle\sum$}}\cr$\scriptstyle\int$\cr}}
	{\ooalign{\raisebox{.2\height}{\scalebox{.6}{$\scriptstyle\sum$}}\cr$\scriptstyle\int$\cr}}
}

\font\ec=ecrm0800 at 12pt
\def\th{\hbox{\ec\char'336}}
\def\edth{\hbox{\ec\char'360}}

\newcommand{\GHPLie}{\text{\L}}
\newcommand{\lie}[1]{\mathcal{L}_#1}

\DeclareMathOperator{\thorn}{\text{\rm \th}}

\def\GHPwt{\ensuremath{\circeq}}
\newcommand{\GHPw}[2]{\left\{ #1, #2 \right\}}

\usepackage{xcolor}
\begin{document}
	
	\title{Canonical Quantization of Teukolsky fields on Kerr Background}
	\author{Claudio Iuliano}%
	\email{iuliano@mis.mpg.de}
	\affiliation{%
		Institute of Theoretical Physics, Leipzig University, Br{\"u}derstra{\ss}e 16, 04103 Leipzig, Germany
	}%
	\affiliation{%
	Max Planck Institute for Mathematics in Sciences (MiS), Inselstra{\ss}e 22, 04103
	Leipzig, Germany
}

	\author{Jochen Zahn}
	\email{jochen.zahn@itp.uni-leipzig.de}
	\affiliation{%
	Institute of Theoretical Physics, Leipzig University, Br{\"u}derstra{\ss}e 16, 04103 Leipzig, Germany
}%
	\begin{abstract}
	Electromagnetic and gravitational perturbations on Kerr spacetime can be reconstructed from solutions to the Teukolsky equations. We study the canonical quantization of solutions to these equations for any integer spin. Our quantization scheme involves the analysis of the Hertz potential and one of the Newman-Penrose scalars, which must be related via the Teukolsky-Starobinsky identities. We show that the canonical commutation relations between the fields can be implemented if and only if the Teukolsky-Starobinsky constants are positive, which is the case both for gravitational perturbations and Maxwell fields. We also obtain the Hadamard parametrix of the Teukolsky equation, which is the basic ingredient for a local and covariant renormalization scheme for non-linear observables. We also discuss the relation of the canonical energy of Teukolsky fields to that of gravitational perturbations.
	\end{abstract}

	\maketitle

	\section{Introduction}
\label{sec:level1}

Quantum field theory on black hole spacetimes is not only crucial for describing black hole evaporation \cite{Zeldovich:1971mw, Unruh:1974bw, Hawking:1975vcx} or for studying quantum effects at the inner (Cauchy) horizon \cite{Zilberman:2019buh, Hollands:2019whz, Klein:2021ctt, Zilberman2022}, related to strong cosmic censorship \cite{Penrose}. Besides those and many other established applications, one could also ask for 
instance whether quantum effects may be used to ``overspin'' an extremal Kerr black hole -- this is classically impossible \cite{Sorce2017}. A natural way to 
investigate this question would be to quantize gravitational perturbations of the extremal Kerr spacetime and then compute semiclassical corrections to its mass and angular momentum.
The aim of this paper is to take a first step towards this difficult problem.

While scalar quantum fields on black hole spacetimes are well understood conceptually and essentially all relevant observables are computable,\footnote{To the best of our knowledge, an explicit computation of the expectation value of the stress tensor in the exterior region of the Kerr spacetime has not yet been performed, but in view of the recent rapid progress (with results for the interior region \cite{Zilberman2022}), this seems to be just a matter of time.} the treatment of gravitational perturbations, but also of Maxwell fields, is less well developed. One difficulty with these is that they are gauge theories, a further one that their field equations are not separable in black hole spacetimes.

Candelas, Chrzanowski and Howard (CCH) \cite{Candelas1981} have suggested an approach to overcome some of the technical difficulties with gravitational perturbations (and analogously for Maxwell fields) on Kerr spacetime by expressing them in terms of the corresponding complex Hertz potential. This is a solution to one of the Teukolsky equations (TE) \cite{Teukolsky1973}, which is well-known to be separable. Hence, one can construct mode solutions for the Hertz potentials. In the CCH approach, these are symplectically normalized by reconstructing the corresponding metric perturbation and using the symplectic inner product that is naturally defined for these. In this way, one obtains quantum fields fulfilling canonical commutation relations (CCR), and one can easily define a state in the usual way, such as the Boulware vacuum. From the metric perturbation, one can also construct the gauge invariant Newman-Penrose (NP) scalars, which can thus be expressed in terms of modes and creation/annihilation operators, so that computations of expectation values (or differences thereof) are possible (at least in certain limits) \cite{Candelas1981,Jensen1995, Casals2005}.

While the CCH approach is very appealing, it has some aspects which are not completely satisfactory or where a deeper understanding seems desirable. A slightly awkward aspect is that the two sets of mode solution of the TE, the $\ri$- and $\ru$- modes (see Sect. \ref{modeexpansion}), are reconstructed differently, i.e., the corresponding metric perturbations are in different gauges (so that the full metric perturbation is not in a well-defined gauge). In the computations that are performed in \cite{Candelas1981,Jensen1995, Casals2005} this is irrelevant, as only the gauge invariant NP scalars are considered. However, to the best of our knowledge, no computation of renormalized expectation values has yet been performed for the NP scalars (only differences of expectation values in different states). To perform a proper renormalization (for example of the stress tensor of the Maxwell field) according to the principles of quantum field theory on curved spacetimes (QFTCS) \cite{WaldQFTCS, Hollands2015} a Hadamard parametrix is necessary. One could of course obtain one by first setting up a parametrix for the gravitational perturbations (for which a choice of gauge would be necessary) and then acting on it with the appropriate differential operators (mapping a metric perturbation to an NP scalar). But to the best of our knowledge, this cumbersome procedure has not yet been performed.

The variation of the CCH approach that we propose here overcomes these difficulties. It is based on the insight that the Hertz potential $\phi$ can naturally be interpreted as ``dual'' to an NP scalar $\psi$ (${}_{2}\psi$ for metric perturbations and ${}_{1}\psi$ for the Maxwell field) in the sense that both the TE for the Hertz potential $\phi$ and the NP scalar $\psi$ follow from the same ``Teukolsky action'', in which $\phi$ and $\psi$ are coupled \cite{Toth2018}. This coupling of $\phi$ and $\psi$ is analogous to the coupling of a charged scalar $\chi$ to its complex conjugate $\chi^*$. In the latter case, one typically first considers $\chi$ and $\chi^*$ as independent (for the derivation of the equation of motion and symplectic normalization, for example), but in the end one has to make sure that the Hermitean conjugate of $\chi$ coincides with $\chi^*$. This leads to a condition on the normalization of modes which fixes it up to a phase. Similarly, in the present case, we require that $\psi$ is the NP scalar for the metric perturbation (or Maxwell field) reconstructed from the Hertz potential $\phi$. This leads to a relation between $\phi$ and $\psi$ involving a differential operator of order $2 s$ (with $s$ being the spin of the field considered, i.e., $s=1$ for the Maxwell field and $s=2$ for gravitational perturbations), which for $s = 0$ reduces to the relation for the scalar field discussed above. A further analogy with the complex scalar field is that the TE for $\phi$ (and $\psi$) can naturally be interpreted as those of a charged Klein-Gordon field in a complex external potential.

From the ``Teukolsky action'' one directly obtains a symplectic form for the fields $(\phi, \psi)$ and imposing symplectic normalization as well as the consistency relation between $\phi$ and $\psi$ discussed above, one recovers the symplectic normalization used in the CCH approach. One slight advantage of our approach is that at least one of the NP scalars is directly available as a quantum field, i.e., no further differentiations are necessary. Furthermore, from the form of the ``Teukolsky action'' it follows that in physically reasonable (``Hadamard'') states the two point function $\langle \phi(x) \psi(x') \rangle$ has a universal short distance singularity, which is captured by the Hadamard parametrix for the ``Teukolsky operator'' occurring in the TEs. This can be straightforwardly obtained by adapting results for the charged Klein-Gordon field in an external potential \cite{Balakumar2020}. Hence, our approach quite directly yields a parametrix which can be used to subtract short distance singularities in order to obtain renormalized expectation values of electromagnetic and gravitational observables.

Finally, a further advantage of our approach is that associated to the ``Teukolsky action'' there is also a canonical energy for $(\phi, \psi)$, which, up to ``boundary'' terms, coincides with the canonical energy for metric perturbations (or Maxwell fields). The latter quantity is relevant for semiclassical corrections to the black hole mass (and thus also the quantum stability or instability of Kerr spacetime). However, for reason explained in more detail below, its computation seems to be a daunting task, while a computation of the ``Teukolsky canonical energy'' might soon be within reach.

The article is structured as follows: In the next section, we first recall basic concepts, such as Kerr geometry and the Geroch, Held and Penrose (GHP) \cite{Geroch1973} formalism. We also recall the reconstruction of gravitational perturbations (and Maxwell fields) from the Hertz potential and introduce the ``Teukolsky action'' and the corresponding symplectic form. In Section~\ref{quantization}, we quantize by first performing a mode expansion, then imposing symplectic normalization and finally implementing the relation between $\phi$ and $\psi$ discussed above. In that context we also discuss the (impossibility of the) generalization to generic spin, and the relation to the CCH approach. In Section~\ref{pointsplit}, we discuss the Hadamard parametrix for the Teukolsky fields, and in Section~\ref{canonicalenergy} we perform some first steps towards the evaluation of the ``Teukolsky canonical energy''. We conclude with a summary and an outlook.

\section{Setup} \label{setup}
\subsection{Kerr Geometry}
The Kerr metric in Boyer-Lindquist coordinates is
\begin{eqnarray}
g &=& \left( 1 - \frac{2 M r}{\Sigma} \right) \mathrm d t^2 + \frac{4 a r M \sin^2 \theta}{\Sigma} \mathrm d t \mathrm d \varphi +\nonumber\\ & & \hspace{1.5cm}- \frac{\Sigma}{\Delta} \mathrm d r^2 - \Sigma \mathrm d \theta^2- \frac{\Gamma}{\Sigma} \sin^2 \theta \mathrm d \varphi^2
 \label{kerrg}
\end{eqnarray}
with
	\begin{eqnarray}
	\Sigma  &=& r^2 + a^2 \cos^2 \theta,\nonumber 
\\
\Delta  &=& r^2 - 2 M r + a^2,\nonumber
\\
\Gamma  &=& (r^2 + a^2)^2 - a^2 \Delta \sin^2 \theta.
\end{eqnarray}
Here $M\geq0$ and $0\leq a\leq M$ represent the black hole mass and the angular velocity parameter. The function $\Delta$ has two distinct real zeros in $r_\pm=M\pm\sqrt{M^2-a^2}$ when $a\neq M$; the root $r_+$ represents the outer (event) horizon while $r_-$ is the inner (Cauchy) horizon of the black hole. The surface gravity on the event horizon is
\begin{equation}
\kappa=\frac{r_+-r_-}{2(r^2_++a^2)},
\end{equation}
which vanishes in the extremal case $a=M$, i.e., when the roots of $\Delta$ coincide.

In our calculations, we will use the tortoise coordinate $r^*$ implicitly defined by
\begin{equation}
\frac{\ud r^*}{\ud r}=\frac{r^2+a^2}{\Delta}.
\end{equation}
From the explicit form of the metric (\ref{kerrg}), this spacetime is stationary and axisymmetric with two Killing vectors:
\begin{equation}
\left(\partial_t\right)^a, \qquad \left(\partial_\varphi\right)^a.
\end{equation} 
The Kerr metric is of Petrov type D and hence possesses two principal null directions $l$ and $n$, i.e., null vector fields such that
\begin{equation}
C_{abc[d}l_{e]} l^a l^b=0, \qquad C_{abc[d}n_{e]} n^a n^b=0,
\end{equation}
with $C_{abcd}$ the Weyl tensor (which coincides with the Riemann tensor on Kerr spacetime).

\subsection{GHP Formalism}\label{GHPform}

In order to simplify Maxwell/linearised Einstein equations around a Kerr background $(\mathcal M, g)$, we use the framework introduced by Geroch, Held and Penrose \cite{Geroch1973}, which is a powerful tool in classical black hole perturbation theory.
One first completes the null directions $n^a, l^a$ to a complex null tetrad $\{l,n,m,\bar m\}$ normalized as
\begin{equation}
l^a n_a=1, \qquad m^a\bar m_a=-1,
\end{equation}
and the other contractions vanishing. Using these, the metric can be written as
\begin{equation}
g_{ab}=2l_{(a} n_{b)}-2m_{(a}\bar m_{b)}.
\end{equation}

The GHP formalism emphasizes the notions of spin and boost weights, defined as follows. The Abelian subgroup of the (local) Lorentz group which preserves the principal null directions $l^a$, $n^a$ and the orthogonality relations is defined by
\begin{equation}
l^a\mapsto\lambda\bar\lambda l^a, \quad n^a\mapsto \left(\lambda\bar\lambda\right)^{-1} n^a, \quad m^a\mapsto\lambda \bar\lambda^{-1}m^a \label{tetradtransfo}
\end{equation} 
with  $\lambda:\mathcal M\to\mathbb C^\times$ and $\mathbb C^\times$ the multiplicative group of complex numbers.
A scalar $\eta$ has GHP weights $\{p,q\}$ if it transforms as
\begin{equation}
\eta\mapsto\lambda^p\bar\lambda^q\eta
\end{equation}
under \eqref{tetradtransfo} and we will write $\eta\GHPwt\{p,q\}$.
For any scalar of type $\{p,q\}$ we can define the spin and the boost weights by
\begin{equation}
s=\frac{p-q}{2},\qquad b=\frac{p+q}{2}.
\end{equation}
Only objects with same weights can be added together and multiplication between $\{p,q\}$ and $\{p',q'\}$ scalars, gives a $\{p+p',q+q'\}$ scalar. The generalization to tensors with GHP weights is straightforward: a tensor $T^{a_1,\dots,a_k}_{b_1,\dots,b_k}$ has GHP weights $\{p,q\}$ if it transforms as
\begin{equation}
	T^{a_1,\dots,a_k}_{b_1,\dots,b_k}\to \lambda^{p}\bar \lambda^{q} T^{a_1,\dots,a_k}_{b_1,\dots,b_k}
\end{equation} 
under tetrad transformations \eqref{tetradtransfo}, and it satisfies the standard transformation law for tensors under change of coordinates.

In this tetrad formalism, there exist $2$ different discrete transformations that reflect the inherent symmetries of this construction:
\begin{itemize}
	\item $'$ : $l^a\leftrightarrow n^a$ and $m^a\leftrightarrow \bar m^a$, $\{p,q\}\mapsto\{-p,-q\}$;
	\item  $\bar{} $ : $m^a\leftrightarrow\bar m^a$, $\{p,q\}\mapsto\{q,p\}$ (Complex conjugation).
\end{itemize}
By taking the directional derivative of the tetrad vectors, the 12 spin coefficients can be defined
\begin{align}
\kappa&=l^{a} m^{b} \nabla_{a} l_{b}, & \sigma&=m^{a} m^{b} \nabla_{a} l_{b},\nonumber \\ \rho&=\bar{m}^{a} m^{b} \nabla_{a} l_{b}, & \tau&=n^{a} m^{b} \nabla_{a} l_{b}
\end{align}
and
\begin{eqnarray}
\beta&=&-\frac{1}{2}\left(m^{a} \bar{m}^{b} \nabla_{a} m_{b}-m^{a} n^{b} \nabla_{a} l_{b}\right),\nonumber\\ \varepsilon&=&-\frac{1}{2}\left(l^{a} \bar{m}^{b} \nabla_{a} m_{b}-l^{a} n^{b} \nabla_{a} l_{b}\right) \label{noghpcovariant}
\end{eqnarray}
with their primed, complex conjugated and prime-complex conjugated versions. Observe that (\ref{noghpcovariant}) do not have a well-defined GHP weight, but they can be encoded in the Lie$(\mathbb C^\times)$ connection
\begin{equation}
\omega_a=\varepsilon n_a-\varepsilon^{\prime} l_a+\beta^{\prime} m_a-\beta \bar{m}_a
\end{equation}
which transforms precisely as a connection one-form $\omega_a\to\omega_a+\lambda^{-1}\nabla_a \lambda$. 
The GHP covariant derivative is
\begin{equation}
\Theta_a=\nabla_a-p\omega_a-q\bar\omega_a,
\end{equation}
 which reduces to the standard covariant derivative when applied on GHP tensors of type $\{0,0\}$.
The projections of the GHP covariant derivative along the tetrad legs are usually called
\begin{eqnarray}
	\th=l^a\Theta_a, \ \ \th'=n^a\Theta_a, \ \ \edth=m^a\Theta_a, \ \ \edth'=\bar m^a \Theta_a.
\end{eqnarray}
When applied to a GHP tensor of type $\{p,q\}$, they give new GHP tensors of type $\{p+p',q+q'\}$ with $\{p',q'\}$ given by
\begin{eqnarray}
\th&\GHPwt&\{1,1\}, \quad \th'\GHPwt\{-1,-1\}, \\ \edth&\GHPwt&\{1,-1\}, \quad \edth'\GHPwt\{-1,1\}.
\end{eqnarray}

Observe that the GHP covariant derivative can be rewritten as
\begin{equation}
\Theta_{a}=l_{a} \th'+n_{a} \th-m_{a} \edth'-\bar{m}_{a} \edth\GHPwt\{0,0\},\label{thetadef}
\end{equation}
which manifestly shows that this operator is invariant under the tetrad transformation (\ref{tetradtransfo}).

On a vacuum solution to the Einstein equation, the non-zero components of the Riemann tensor are given by the components of the Weyl tensor:
\begin{align}
\Psi_{0}&=-C_{l m l m}\GHPwt\{4,0\}, \quad &\Psi_{1}&=-C_{l n l m}\GHPwt\{2,0\},\nonumber \\ \Psi_{2}&=-C_{l m \bar{m} n}\GHPwt\{0,0\}, \quad &\Psi_{3}&=-C_{l n \bar{m} n}\GHPwt\{-2,0\},\nonumber \\  \Psi_{4}&=-C_{n \bar{m} n \bar{m}}\GHPwt\{-4,0\}.
\end{align}
In particular for Kerr geometry, we have further simplifications. Using the two principal null directions $l$ and $n$, one finds 
\begin{align}
&\kappa=\kappa'=\sigma=\sigma'=0 \nonumber\\
&\Psi_0=\Psi_1=\Psi_3=\Psi_4=0, \ \ \Psi_2=-\frac{M}{\zeta^3}, \label{psi2}
\end{align}
with $\zeta:=r-ia\cos\theta$.
Other simplifications read \cite{Pound2021}
\begin{equation}
\frac{\rho}{\bar{\rho}}=\frac{\rho^{\prime}}{\bar{\rho}^{\prime}}=-\frac{\tau}{\bar{\tau}^{\prime}}=-\frac{\tau^{\prime}}{\bar{\tau}}=\frac{\Psi_{2}^{1 / 3}}{\bar{\Psi}_{2}^{1 / 3}}=\frac{\bar{\zeta}}{\zeta}.
\end{equation}

\subsection{Wald identity and Teukolsky action}
\label{sec:WaldIdentity}
On Kerr spacetime, Teukolsky \cite{Teukolsky1973} proved that the NP components of the Maxwell tensor with highest and lowest spin ($s=\pm1$) and the components of the perturbed Weyl tensor with highest and lowest spin ($s=\pm2$) satisfy second order differential equations which are uncoupled and separable. Namely, consider the following NP scalars ${}_{s}\psi\GHPwt\{2s,0\}$ to be
\begin{eqnarray}
{}_{1}\psi&=&F_{l m},\nonumber \\ {}_{2}\psi&=&-R^{(1)}_{lmlm} 
\end{eqnarray} 
where $F$ is the Maxwell tensor and $R^{(1)}$ is the first order correction to the Weyl tensor, then the Teukolsky equations can be written in GHP formalism as
\begin{equation}
{}_{s}\mathcal O{}_{s}\psi=\left[g^{ab}(\Theta_a+2sB_a)(\Theta_b+2sB_b)-4s^2\Psi_2\right]{}_{s}\psi=0, \label{teukol}
\end{equation}
with the null vector $B^a=-\rho n^a+\tau \bar m^a\GHPwt\{0,0\}$. Observe in particular that ${}_{s}\mathcal O$ maps GHP scalars of type $\{p,0\}$ into scalars of type $\{p,0\}$. Moreover, (\ref{teukol}) shows that the TE has the structure of a Klein-Gordon equation in an external potential. Indeed, on any ${}_{s}\eta\GHPwt\GHPw{2s}{0}$,
\begin{eqnarray}
(\Theta_a+ 2sB_a){}_{ s}\eta&=&\left[\nabla_a+ 2s\left(B_a-\omega_a\right)\right]{}_{s}\eta \nonumber\\
&=&(\nabla_a+ s\Gamma_a ){}_{ s}\eta\label{KGrelation} 
\end{eqnarray}
where $\Gamma_a:=2\left[(-\varepsilon-\rho) n_a+\varepsilon' l_a-\beta'm_a+(\beta+\tau) \bar m_a\right]$ can be understood as a (complex) external vector potential.

Based on the adjoint method, Wald proved \cite{Wald78}  important relations between the kernel of the adjoint of ${}_{s} \mathcal O$ and the solutions of the Maxwell operator ($s=1$) and the linearised  Einstein equation ($s=2$). Consider an operator $\mathcal P$ taking an $n-$index tensor field to $m-$index tensor field, we say that $\mathcal P^\dagger$ is the adjoint of $\mathcal P$ if
\begin{equation}
\psi^{a_{1} \ldots a_{m}}(\mathcal{P} \phi)_{a_{1} \ldots a_{m}}-\left(\mathcal{P}^{\dagger} \psi\right)^{a_{1} \ldots a_{n}} \phi_{a_{1} \ldots a_{n}}=\nabla_{a} s^{a} \label{duality}
\end{equation}
with $s^a$ a vector field depending locally on $\psi$ and $\phi$.

We can think of ${}_{s}\psi$ as being obtained by a linear differential operator ${}_{s}\mathcal T$
\begin{equation}
{}_{s}\psi={}_{s}\mathcal T(f),\qquad f=\begin{cases}
h \qquad \text{for $s=2$}\\
A \qquad \text{for $s=1$}
\end{cases}\label{fieldup}
\end{equation}
where $h$ is the first order perturbation of the metric and $A$ is the electromagnetic potential. The operator ${}_{s}\mathcal T$ maps GHP quantities of type $\{0,0\}$ into $\{2s,0\}$. Moreover, let ${}_{s}\mathcal E$ be the field equation for $f$
\begin{equation}
{}_{s}\mathcal E(f)=J, \label{eqforf}
\end{equation}
where $J$ is a source term. From Teukolsky's derivation \cite{Teukolsky1973}, one identifies the differential operator  ${}_{s}\mathcal S$ of order $s$ such that
\begin{equation}
{}_{s}\mathcal S(J)={}_{s}\mathcal O{}_{s}\psi, \qquad {}_{s}\mathcal{S}:\{0,0\}\to\{2s,0\}
\end{equation}
are the inhomogeneous TEs. Then the identity
\begin{equation}
{}_{s}\mathcal S {}_{s}\mathcal E (f)={}_{s}\mathcal O {}_{s}\mathcal T(f) \label{waldid}
\end{equation}
is known as Wald identity.
Finally, if ${}_{s}\mathcal E={}_{s}\mathcal E^\dagger$, which is the case for linearised Einstein and Maxwell equation, and ${}_{-s}\phi$ is a solution to $({}_{s}\mathcal O)^\dagger {}_{-s}\phi=0$, then
\begin{equation}
0={}_{s}\mathcal T^\dagger {}_{s}\mathcal O^\dagger  {}_{-s}\phi={}_{s}\mathcal E {}_{s}\mathcal S^\dagger {}_{-s}\phi.
\end{equation}
In other words, ${}_{s}\mathcal S^\dagger:\ker{}_{s}\mathcal O^\dagger\to\ker{}_{s}\mathcal E$. The field
\begin{equation}
f=\Re[{}_{s}\mathcal S^\dagger {}_{-s}\phi]\label{reco}
\end{equation}
is called reconstructed field and it satisfies ${}_{s}\mathcal E(f)=0$. Using ${}_{s}\mathcal S^\dagger$ defined in \cite{Chrzanowski1975,Cohen1974,Wald78}, the field $f$ derived from \eqref{reco} is in the so-called ingoing radiation gauge, namely
\begin{itemize}
	\item for $s=1$ one has $l^a A_a=0$;
	\item for $s=2$ one has $l^a h_{ab}=0=\mathrm{tr}(h_{ab})$.
\end{itemize} 

 The operator
\begin{equation}
{}_{s}\mathcal O^\dagger=\left[g^{ab}(\Theta_a-2sB_a)(\Theta_b-2sB_b)-4s^2\Psi_2\right]={}_{-s}\mathcal O \label{teukoldag}
\end{equation}
maps GHP scalars of type $\{-p,0\}$ into scalars of type $\{-p,0\}$. An element ${}_{-s}\phi\GHPwt\{-2s,0\}$ in the kernel of ${}_{s}\mathcal O^\dagger$, i.e., fulfilling
\begin{equation}
\label{eq:O_phi}
  {}_{s}\mathcal O^\dagger {}_{-s}\phi = 0,
\end{equation}
is called a Hertz potential. 
According to \eqref{fieldup} and \eqref{reco}, the NP scalar ${}_{s}\psi$ corresponding to the field $f$ reconstructed from the Hertz potential ${}_{-s}\phi$ is
\begin{equation}
{}_{s}\psi={}_{s}\mathcal T\left(\Re[\mathcal {}_{s}\mathcal S^\dagger{}_{-s}\phi]\right) \label{psiphiconstr}.
\end{equation}
To maintain this consistency condition between the Hertz potential ${}_{-s}\phi$ and the NP scalar ${}_{s}\psi$ at the quantum level will be a crucial aspect of our work.

In the following, we consider the pair $\Psi=({}_{-s}\phi,{}_{s}\psi)$ of Teukolsky fields and notice that their equations of motion \eqref{teukol}, \eqref{eq:O_phi} follow from the ``Teukolsky action'' \cite{Toth2018} (recall that $\Psi_2$ was given in \eqref{psi2})
\begin{align}
S[\Psi] = \int \Re \left[ (\Theta^a - 2s B^a) {}_{-s}\phi (\Theta_a + 2s B_a) {}_{s}\psi +\nonumber\right.\\\left. + 4s^2 \Psi_2\; {}_{-s}\phi {}_{s}\psi \right]  \vol_g. \label{teukolaction}
\end{align}
To this action corresponds the symplectic form
\begin{equation}
 \tilde \sigma(\Psi,\Psi')= \int_{\Sigma}\ud\Sigma_a j^a(\Psi,\Psi'),\label{symp}
\end{equation}
with $\Sigma$ a spacelike Cauchy hypersurface, $\ud \Sigma_a$ the corresponding future directed area element, and the symplectic current
\begin{eqnarray}
j^a(\Psi,\Psi')&=&{}_{-s}\phi (\Theta^a+2sB^a){}_{s}\psi' -{}_{-s} \phi'  (\Theta_a+2sB^a){}_{s}\psi \nonumber \\ & &\hspace{-1cm} - {}_{s} \psi' (\Theta^a-2sB^a){}_{-s}\phi +  {}_{s}\psi (\Theta^a-2sB^a){}_{-s}\phi'.
\end{eqnarray}
Using the fact that $\Theta_a$ reduces to $\nabla_a$ on GHP tensors of type $\GHPw{0}{0}$ and $j^a\GHPwt\GHPw{0}{0}$, one easily checks that
\begin{equation}
\nabla_a j^a=\Theta_aj^a=0
\end{equation}
on-shell, i.e., when the components of Teukolsky fields $\Psi$, $\Psi'$ fulfill the equations of motion \eqref{teukol}, \eqref{eq:O_phi}. Hence, the symplectic form $\tilde \sigma$ is independent of the choice of the Cauchy surface $\Sigma$.

The symplectic form \eqref{symp} will be the starting point of our canonical quantization. Note that we still need to impose the constraint \eqref{psiphiconstr}, which also guarantees that we have the correct number of degrees of freedom (two, from the complex Hertz potential).

 \section{Quantization of Teukolsky fields} \label{quantization}
 \subsection{Mode Expansion} \label{modeexpansion}
 We now want to find an explicit representation of the quantum field operators, following the usual procedure starting from a complete set of modes which are normalized w.r.t.\ the symplectic form $\tilde \sigma$. We work in the Kinnersley frame \cite{Kinnersley}
 (given in App.~\ref{kinnframe}), in which the TEs are separable.
 
 We make the usual ansatz
 \begin{eqnarray}
 {}_{-s}v_{\omega,\ell m}&=&{}_{-s}\mathcal N_{\omega,\ell m}e^{-i\omega t}e^{im\varphi}{}_{-s}R(r)_{\omega,\ell m}{}_{-s}S_{\omega,\ell m}(\theta),\nonumber\\ {}_{s}u_{\omega,\ell m}&=&{}_{s}\mathcal N_{\omega,\ell m}e^{-i\omega t}e^{im\varphi}{}_{s}R(r)_{\omega,\ell m}{}_{s}S_{\omega,\ell m}(\theta),
	\label{modexp}
 \end{eqnarray}
 where the radial function $R(r)$ and the angular function $S(\theta)$ satisfy the Teukolsky radial and angular equation \cite{Teukolsky1973}. The coefficients $\mathcal N$ will be defined (up to a phase) in the next sections. In order to uniquely fix the solutions \eqref{modexp} (up to a phase), we need to impose the asymptotic behaviours at the null boundaries of the spacetime. We can choose as a basis of solutions to the radial TEs two classes of solutions known as in- and up-modes defined by their behaviours on the past null infinity $\mathcal I^-$ and past horizon $\mathcal H^-$:
 \begin{itemize}
 	\item in-modes: representing waves coming from $\mathcal I^-$, characterized by
 	\begin{equation}
 	{}_{s}R^{\ri}_{\omega,\ell m}(r)\sim\begin{cases}
 	{}_{s}\mathcal T^{\ri}_{\omega,\ell m}\;\Delta^{-s}e^{-i k r_*} &\, r_*\to-\infty\\
 	\frac{1}{r}e^{-i\omega r_*}+{}_{s}\mathcal R^{\ri}_{\omega,\ell m} r^{-1-2s}e^{i\omega r_*} &\, r_*\to\infty \label{inpositive}
 	\end{cases}
 	\end{equation}
 	\item up-modes: representing waves coming from $\mathcal H^-$, characterized by
 	\begin{equation}
 	{}_{s}R^{\ru}_{\omega,\ell m}(r)\sim\begin{cases}
 	e^{ik r_*}+{}_{s}\mathcal R^{\ru}_{\omega,\ell m}\Delta^{-s}e^{-ik r_*} &\, r_*\to-\infty\\
 	{{}_{s}\mathcal T^{\ru}_{\omega,\ell m}}r^{-1-2s}e^{i\omega r_*} &\, r_*\to\infty.\label{uppositive}
 	\end{cases}
 	\end{equation}
 \end{itemize}
 Here $\mathcal R$ and $\mathcal T$ are the reflection and transmission coefficients, the power-laws $\Delta^{-s}$ and $r^{-1-2s}$ are derived in \cite{Teukolsky1973}, and
 \begin{equation}
 k=\omega-m\omega_+, \ \ \omega_+=\frac{a}{2Mr_+}.
 \end{equation}
 
 With these prescriptions, we define ${}_{s}u^{\ri/\ru}_{\omega,\ell m}$ and ${}_{-s}v^{\ri/\ru}_{\omega,\ell m}$ in accordance with (\ref{modexp}). Observe that if $0<\omega<m\omega_+$ then $k<0$. For these frequencies, the up-modes do not describe incoming waves from the past horizon, but waves to the past horizon.
 These modes are related to the so-called superradiance (see \cite{Ottewill2000} for a discussion in the scalar case). In order to account for this in our mode expansions, we relabel the up-modes $u^{\ru}_{\omega(k),\ell m}\to u^{\ru}_{k,\ell m}$ and use as mode basis
 \begin{equation}
{}_{s}u^{\ri}_{\omega,\ell m} \text{ for $\omega>0$}, \ \ {}_{s}u^{\ru}_{k,\ell m} \text{ for $k>0$},
 \end{equation}
 and analogously for the ${}_{-s}v$ fields.
 
 In Sect. \ref{normalization} and \ref{constraintsect} , we will prove that to fulfil the CCRs and the constraint (\ref{psiphiconstr}), generalized to any integer $s$, the normalization constants must be chosen such that
 \begin{align}
 1 & =2\pi (2\omega)^{2s+1}|{}_{-s}\mathcal N^{\ri}_{\omega,\ell m}|^2 , \\
 1 & =8\pi p_s |{}_{-s}\mathcal N_{k,\ell m}^{\ru}|^2 (k M r_+)^{-1} \nonumber \\
 & \quad \times \prod_{j=1}^{s-1}\lvert 4kMr_++2i(s-j)\sqrt{M^2-a^2}\rvert^{-2}. \label{normconditions}
 \end{align}
In the second expression, $p_s$ is the radial Teukolsky-Starobinsky constant \cite{Chandrasekhar1978,Casals2021} and the product is $1$ for $s=1$. For a consistent quantization, $p_s$ must be positive for all $k, \ell, m$.

Finally, in Sect. \ref{quantum} we will argue that a representation of gravitational and electromagnetic quantum operators, can be obtained by only considering the ones reconstructed from the Hertz potentials by \eqref{reco}.
 
 \subsection{Symplectic normalization} \label{normalization}
From the previous section, we know that a basis on the space of Teukolsky fields $\Psi$ can be written as:
\begin{align}
{}_{-}\Phi^{\ri}_{\omega \ell m}& = \begin{pmatrix} _{-s}v^{\ri}_{\omega \ell m}\vspace{0.1cm} \\ 0 \end{pmatrix}, &\ \
{}_{-}\Phi^{\ru}_{k \ell m} &= \begin{pmatrix} _{-s}v^{\ru}_{k \ell m}\vspace{0.1cm} \\ 0 \end{pmatrix},\nonumber \\
{}_{+}\Phi^{\ri}_{\omega \ell m}& = \begin{pmatrix} 0\vspace{0.1cm} \\ _{s}u^{\ri}_{\omega \ell m} \end{pmatrix}, &\ \
{}_{+}\Phi^{\ru}_{k \ell m}& = \begin{pmatrix} 0\vspace{0.1cm} \\ _{s}u^{\ru}_{k \ell m} \end{pmatrix}.
\end{align}
Since the symplectic form is conserved, i.e., independent of $\Sigma$, we will evaluate it in the limit $\Sigma\to \mathcal H^{-}\cup \mathcal I^-$. In particular, we consider the hypersurface $\Sigma$ at $t=t_0$ and we will take the limit $t_0\to-\infty$. The volume element on $\Sigma$ is $\ud \Sigma_a = u_a \sqrt{-q} \ud r \ud \theta \ud \phi$ with the future pointing unit normal
\begin{equation}
u_a=\sqrt{\frac{\Sigma\Delta}{\Gamma}}(1,0,0,0),\quad u^a=\left(\sqrt{\frac{\Gamma}{\Sigma \Delta}},0,0,\frac{2Mra}{\sqrt{\Sigma\Delta\Gamma}}\right),\label{normal}
\end{equation}
and the determinant of induced metric $q_{ab}$
\begin{equation}
q=-\frac{\Sigma\Gamma}{\Delta}\sin^2\theta.
\end{equation}

In the limit $t_0\to -\infty$, the in- and up- contributions to the symplectic form decouple and, noticing that the symplectic form $\tilde \sigma$ mixes opposite spin fields, we impose the  normalization conditions
\begin{align}
\tilde \sigma \left( {}_{\mp}\Phi^{\ri}_{\omega \ell m},{}_{\pm}\Phi^{\ri}_{-\omega' \ell' -m'}\right)&= i\delta(\omega-\omega')\delta_{\ell \ell'}\delta_{mm'}, \nonumber\\
\tilde \sigma \left( {}_{\mp}\Phi^{\ru}_{k \ell m},{}_{\pm}\Phi^{\ru}_{-k' \ell' -m'}\right)&= i\delta(k-k')\delta_{\ell \ell'}\delta_{mm'}. \label{sympnorm}
\end{align}
 
 Let us evaluate these conditions starting with the in-modes. In the Kinnersley frame, the external vector potential $\Gamma^a$ defined below \eqref{KGrelation} reads:
 \begin{eqnarray}
 \Gamma^t & = & - \frac{1}{\Sigma} \left[ \frac{M (r^2 - a^2)}{\Delta} - (r + i a \cos \theta) \right],\nonumber \\
 \Gamma^r & =& - \frac{1}{\Sigma} (r - M),\nonumber \\
 \Gamma^\theta & =& 0, \nonumber\\
 \Gamma^\varphi & = &- \frac{1}{\Sigma} \left[ \frac{a (r-M)}{\Delta} + i \frac{\cos \theta}{\sin^2 \theta} \right]. \label{gaugepot}
 \end{eqnarray}
 On $\mathcal H^-$ the in-modes vanish and we have to perform the integral in \eqref{symp} only on the past null infinity. On $\mathcal{I}^-$, all the $\Gamma$ components vanish, and using \cite{Pound2021}
 \begin{equation}
 \label{eq:S_S}
 	{}_{s}S_{\omega,\ell m}(\theta)=(-1)^{m+s}{}_{-s} S_{-\omega,\ell -m}(\theta)
 \end{equation} 
 together with  the orthogonality of spheroidal harmonics,
 \begin{equation}
 \int_0^\pi \ud \theta \sin\theta {}_{\pm s} S_{\omega,\ell m} \,{}_{\pm s}S_{\omega, \ell' m}=\delta_{\ell \ell'},
 \end{equation}
 we find
 \begin{equation}
 \begin{split}
\tilde \sigma( &{}_{\mp}\Phi^{\ri}_{\omega \ell m}, {}_{\pm}\Phi^{\ri}_{- \omega' \ell' - m'}) =\\ &(-1)^{m+s}    {}_{\mp s}\mathcal N^\ri_{\omega \ell m} \ _{\pm s}\mathcal N^\ri_{- \omega \ell -m} 4i \pi \omega\, \delta(\omega-\omega')\delta_{mm'}\delta_{\ell \ell'}. \label{innorm}
\end{split}
\end{equation}
  In order to obtain the symplectic normalization \eqref{sympnorm}, we must have
 \begin{equation}
 \label{eq:SymplecticModeNormalization}
 {}_{\pm s}\mathcal N^{\ri}_{\omega \ell m} \ _{\mp s}\mathcal N^{\ri}_{-\omega \ell -m} = \frac{(-1)^{m+s}}{4 \pi \omega}.
 \end{equation}
 
 For the up-modes ${}_{\pm}\Phi^{\ru}_{k,lm}$, we perform the integral in \eqref{symp} only on $\mathcal H^-$ since on $\mathcal I^-$ these modes vanish. In this case, the components $\Gamma^a$ do not vanish on $\mathcal H^-$. 
 We have to compute the contraction
 \begin{equation}
 u^a\Theta_a=u^a\left(l_{a} \thorn'+n_{a} \thorn-m_{a} \edth'-\bar{m}_{a}\edth \right).
 \end{equation}
 Since we want to compute the symplectic form on the Cauchy surface $t_0\to-\infty$ on the field $\Phi^{\ru}$, the only contributions come from $r\to r_+$, i.e. $\Delta\to0$. In this limit, we observe that the contributions arise in the contractions with the legs $l_a$ and $n_a$: indeed one can easily see from the form of \eqref{normal} that when $\Delta\to0$ only those legs and those GHP operators which contain factors of the form $1/\sqrt{\Delta}$ contribute to the integral. Moreover, making use of \eqref{KGrelation}, we find that for $r\to r_+$
 \begin{equation}
 u^a(\nabla_a\pm s \Gamma_a)\simeq\frac{1}{\sqrt{\Sigma\Delta}}\left[(r^2_++a^2) \partial_t+a\partial_\phi\mp s(r_+-M)\right].
 \end{equation}
 Thus, on the past horizon, we get
 \begin{equation}
 \begin{split}
 \tilde \sigma(& {}_{\mp}\Phi^{\ru}_{k \ell m}, {}_{\pm}\Phi^{\ru}_{- k' \ell' - m'} )
= (-1)^{m+s} 2 \pi{}_{\mp } \mathcal{N}_{k \ell m}^{\ru} {}_{\pm } \mathcal{N}_{-k \ell-m}^{\ru} \\& \times\left(4 i k M r_{+}\mp 2 s \sqrt{M^2-a^2}\right)   \delta\left(k-k^{\prime}\right)\delta_{m m^{\prime}}\delta_{\ell \ell^{\prime}}.
  \end{split}
 \end{equation}
In order to have the correct symplectic normalization \eqref{sympnorm}, we impose
 \begin{equation}
 {}_{\mp s}\mathcal N^{\ru}_{k,\ell m}{}_{\pm s}\mathcal{N}^{\ru}_{-k,\ell-m}=\frac{(-1)^{m+s}}{2\pi(4kMr_+\pm i 2s\sqrt{M^2-a^2})}. \label{upnorm}
 \end{equation}

 As one can easily see, \eqref{innorm} and \eqref{upnorm} do not yet fix the normalization coefficients uniquely (not even up to a phase). We have the freedom to modify them using a complex $\lambda$ by
 \begin{equation}
 \begin{split}
 \label{eq:NormalizationChange}
 &_{\pm s}\mathcal N^{\ri}_{\omega \ell m}  \to \lambda_{\omega \ell m}^{\ri} \ _{\pm s}\mathcal N^{\ri}_{\omega \ell m}, \\
 &_{\mp s}\mathcal N^{\ri}_{-\omega \ell - m} \to (\lambda^{\ri}_{\omega \ell m})^{-1} \ _{\mp s}\mathcal N^{\ri}_{-\omega \ell -m},
 \end{split}
 \end{equation}
 together with similar transformations for the normalization coefficients for the up-modes. How to fix this ambiguity (up to a phase) will be discussed below.

 The field operators can be expressed using the basis described above:
  \begin{equation}
 \begin{split}
 {}_{s}\psi=& \sum_{\ell, m}\left[\int_{0}^{\infty} \ud \omega\left(a_{\omega \ell m}^{\ri} {}_{s}u_{\omega \ell m}^{\ri }+b_{\omega \ell m}^{\ri\dagger} {}_{s} u_{-\omega \ell -m}^{\ri }\right)+\right.\\&\left.+\int_{0}^{\infty} \ud k\left(a_{k \ell m}^{\ru} {}_{s}u_{k \ell m}^{\ru }+b_{k \ell m}^{\ru \dagger} {}_{s} u_{-k \ell -m}^{\ru}\right)\right]\\
 {}_{-s}\phi=& \sum_{\ell, m}\left[\int_{0}^{\infty} \ud \omega\left(b_{\omega \ell m}^{\ri} {}_{-s}v_{\omega \ell m}^{\ri }+a_{\omega \ell m}^{\ri \dagger} {}_{-s} v_{-\omega \ell -m}^{\ri }\right)+\right.\\&\left.+\int_{0}^{\infty} \ud k\left(b_{k \ell m}^{\ru} {}_{-s}v_{k \ell m}^{\ru}+a_{k \ell m}^{\ru \dagger} {}_{-s} v_{-k \ell -m}^{\ru }\right)\right].
 \end{split} \label{expansion}
 \end{equation}

 Because we have normalized the modes using the symplectic form $\tilde \sigma$, the creation and annihilation operators must fulfil
 \begin{equation}
 \begin{split}
 [a^{\ri}_{\omega,\ell m},a^{\ri,\dagger}_{\omega',\ell'm'}]&=\delta(\omega-\omega')\delta_{\ell\ell'}\delta_{mm'}, \\
 [a^{\ru}_{k,\ell m},a^{\ru,\dagger}_{k',\ell'm'}]&=\delta(k-k')\delta_{\ell\ell'}\delta_{mm'},
 \end{split}
 \end{equation}
 and analogously for $b^{\ri/\ru}$ (the others commutators vanish) in order for the Teukolsky fields to fulfil the CCR. We define the past Boulware vacuum state $|B\rangle$ by
 \begin{equation}
 \begin{aligned}
 a_{\omega \ell m}^{\ri}\left|B\right\rangle &=0, & b_{\omega \ell m}^{\ri}\left|B\right\rangle=0, \\
 a_{k \ell m}^{\ru}\left|B\right\rangle &=0, & b_{k \ell m}^{\ru}\left|B\right\rangle=0,
 \end{aligned}\label{paststate}
 \end{equation}
 corresponding to an absence of particles emerging from $\mathcal H^-$ and $\mathcal I^-$ \cite{Ottewill2000}. With the previous definitions, we write the
 two-point function of the field in the past Boulware vacuum states
 \begin{align}
 w^{\phi\psi}_B(x,x') & := \langle B| {}_{-s}\phi(x) {}_{s}\psi(x')|B\rangle \nonumber \\
 & = \sum_I\SumInt_\lambda v^I_{\lambda}(x)u^I_{-\lambda}(x') \label{state}
 \end{align}
 where, in order to simplify the notation, we have introduced the indices $I\in\{\ri,\ru\}$, $\lambda=\{\omega_I,\ell m\}$ and $-\lambda=\{-\omega_I,\ell-m\}$, with $\omega_{\ri}=\omega,\, \omega_{\ru}=k$, and $\SumInt$ stands for summation over $\ell$ and $m$ and integration over $\omega_I$ from $0$ to $\infty$. We remind the reader that, due to the limited applications of the Boulware state in the sub-extremal regime of a Kerr spacetime, in Sect. \ref{unruhstate}, we will give a formal construction of Unruh state.
 
 In any case, we conclude by stressing the fact that while the mode expansion is a standard procedure to construct ``vacuum'' states \cite{Ottewill2000}, it does not guarantee the Hadamard property of such states (due to the fact that there is no Killing field which is time-like in the full exterior region). For example, the existing proofs of the Hadamard property of the Unruh state on Kerr (-deSitter) spacetime require small rotation parameter $a$ \cite{Klein2023, Gerard2020}.

 \subsection{Implementation of \eqref{psiphiconstr}} \label{constraintsect}
Finally, we need to implement the relation \eqref{psiphiconstr} between $_{-s}\phi$ and $_{s}\psi$ at the quantum level. In the Maxwell $(s= 1)$ and linearised Einstein theories $(s=2)$, these conditions can be written as the Teukolsky-Starobinsky identity (see e.g. \cite{Ori2003, Pound2021, Casals2021})
\begin{equation}
 \label{eq:RealityCondition}
 _{s}\psi = \frac{(i)^{2s}}{2s} \thorn^{2s}\, _{-s} \bar{\phi},
\end{equation}
but we can (and will) also consider this relation for general spin $s$.\footnote{One can see that in the scalar case $s = 0$ the relation \eqref{psiphiconstr} reduces to the relation between the complex scalar field and its Hermitean conjugate, discussed in the Introduction.}
When expanding the fields as in \eqref{expansion}, the condition \eqref{eq:RealityCondition} is fulfilled iff the modes satisfy
\begin{equation}
\label{eq:ModeConstraint}
 \frac{i^{2s}}{2s}\thorn^{2s} {}_{-s}\overline v^{\ri}_{\omega,\ell m} = {}_{+s} u^{\ri}_{-\omega,\ell-m},
\end{equation}
and analogously for the $\ru$-modes. 

We discuss the gravitational case ($s=2$) first. As we know that complex conjugation followed by application of $\frac{1}{4} \thorn^4$ maps solutions to the TE of spin $-2$ to those of spin $+2$, this must also be true for mode solutions, for which, by the complex conjugation, $\omega$ and $m$ change sign. It can be proved along the lines of \cite{Ori2003} that the mode solution ${}_{-2}v_{\omega,\ell m}$ satisfying both the TE and the constraint \eqref{eq:ModeConstraint}, can be obtained by
 \begin{equation}
 {}_{-2}\bar v_{\omega,\ell m}=p^{-1} \Delta^{2}\left(D^{\dagger}\right)^{4}\left[\Delta^{2} \;{}_{2} u_{-\omega,\ell-m}\right]=: H({}_{2} u_{-\omega,\ell-m}) \label{reconv}
 \end{equation}
 with
 \begin{equation}
 D^{\dagger}=-\frac{r^{2}+a^{2}}{\Delta} \partial_{t}+\partial_{r}-(a / \Delta) \partial_{\varphi}.
 \end{equation}
 In \eqref{reconv}, $p$ is related to the non-vanishing radial Teukolsky-Starobinsky constant \cite{Casals2021}
 \begin{equation}
 \begin{split}
 p=&( \lambda^{2}(\lambda+2)^{2}-8 \omega^{2} \lambda\left[\alpha^{2}(5 \lambda+6)-12 a^{2}\right] \\&
 +144 \omega^{4} \alpha^{4}+144 \omega^{2} M^{2})/4
 \end{split}
 \end{equation}
 where $\alpha=a^2-am/\omega$ and $\lambda$ is defined in \cite{Chandrasekhar1978}. Observe that using Teukolsky-Starobinsky identities, it is possible to see that $p>0$ \cite{Chandrasekhar1978}. 
Hence, we can instead of \eqref{eq:ModeConstraint} equivalently require that
\begin{equation}
  H{}_{+2}u^{\ri}_{-\omega,\ell-m}={}_{-2}\overline v^{\ri}_{\omega,\ell m}, \label{constraints}
\end{equation}
and analogously for the $\ru$-modes. 
 
 In order to explicitly perform  the $r-$derivatives, we work in the asymptotic regions $\mathcal I^-$ and $\mathcal H^-$, where the radial functions have explicit expressions \eqref{inpositive} and \eqref{uppositive}.
 It is important to notice that $\thorn$ vanishes on $e^{i\omega (t-r_*)}$ for $r_*\to\infty$ and on $e^{i\omega t-i k r_*}$ for $r_*\to-\infty$, while $D^\dagger$  vanishes on $e^{i\omega (t+r_*)}$ for $r_*\to\infty$ and on $e^{i\omega t+i k r_*}$ for $r_*\to-\infty$. This leads to a complication in the computations since these will require the analysis of sub-leading terms in the asymptotic expansions of the mode solutions. To overcome this problem, we use the same strategy as in \cite{Ori2003}.  Define:
 \begin{equation}
 D_{\omega,m}=\partial_r+iK/\Delta,\ \ D^\dagger_{\omega,m}=\partial_r-iK/\Delta,
 \end{equation}
 with $ K=am-(r^2+a^2)\omega$. Using \eqref{eq:S_S}, we can express the constraints \eqref{eq:ModeConstraint}, \eqref{constraints} as
\begin{align}
 \label{stepconstraint_in}
 & (-1)^m {}_{2} \mathcal N^{\ri}_{-\omega,\ell-m}{}_{2}R^{\ri}_{-\omega,\ell-m} \\ 
 & \quad ={}_{-2}\overline{\mathcal N}^{\ri}_{\omega,\ell m}\frac{1}{4}(D^\dagger_{\omega,m})^4 {}_{-2}\overline{R}^{\ri}_{\omega,\ell m}, \nonumber \\
 \label{stepconstraint_up}
 & (-1)^m {}_{-2} \overline{\mathcal N}^{\ru}_{\omega,\ell m} {}_{-2}\overline{R}^{\ru}_{\omega,\ell m} \\
 & \quad ={}_{2}\mathcal N^{\ru}_{-\omega,\ell -m} p^{-1}\Delta^2 (D_{\omega,m})^4\left[\Delta^2{}_{2}R^{\ru}_{-\omega,\ell -m}\right]. \nonumber
\end{align}
We have \cite{Ori2003}, for $r_* \to \infty$
\begin{align}
 (D^\dagger_{\omega,m})^4\left(\frac{e^{i\omega r_*}}{r}\right) & \sim 16\omega^4 \left(\frac{e^{i\omega r_*}}{r}\right), \\
 \Delta^2 (D_{\omega,m})^4\left[ \Delta^2\left(\frac{e^{-i\omega r_*}}{r^5}\right)\right] & \sim 16\omega^4 r^3 e^{-i\omega r_*},
\end{align}
 and for $r_* \to - \infty$
\begin{align}
 (D^\dagger_{\omega,m})^4 \left(\Delta^2 e^{ikr_*}\right) & \sim \bar Q \Delta^{-2} e^{ikr_*}, \\
 \Delta^2(D_{\omega,m})^4\left[\Delta^2(e^{-ikr_*})\right] & \sim Q e^{-ikr_*},
 \end{align}
 where
 \begin{equation}
 \begin{split}
 Q=&(4kMr_+ + 4 i\sqrt{M^2-a^2})(4kMr_+ + 2 i\sqrt{M^2-a^2})\times\\&\times(4kMr_+ - 2 i\sqrt{M^2-a^2})4kMr_+.
  \end{split}
 \end{equation}
Using the above to compare the asymptotic behaviour as $r_* \to \infty$ on both sides of \eqref{stepconstraint_in}, we see that it reduces to
\begin{equation}
 (-1)^m {}_{2} \mathcal N^{\ri}_{-\omega,\ell-m} = 4 \omega^4 {}_{-2}\overline{\mathcal N}^{\ri}_{\omega,\ell m}.
\end{equation}
Taking into account the constraint \eqref{eq:SymplecticModeNormalization} from symplectic normalization, we obtain
 \begin{equation}
  16\pi\omega^5 |{}_{-2}\mathcal N^{\ri}_{\omega,\ell m}|^2=1
 \end{equation}
 which can be easily satisfied. Comparing both sides of \eqref{stepconstraint_up} for $r_* \to - \infty$, one similarly obtains, using \eqref{upnorm}
\begin{equation}
 \begin{split}
 \frac{|{}_{-2}\mathcal N^{\ru}_{k,\ell m}|^2 p \pi }{2kMr_+|4kMr_++2i\sqrt{M^2-a^2}|^2}=1,
 \end{split}
 \end{equation}
 which can also be satisfied, as $p$ is positive, see below.

For generic spin $s$, the constraint \eqref{eq:ModeConstraint} can be inverted as
 \begin{equation}
 {}_{-s}\bar v_{\omega,\ell m}= i^{2s} p_s^{-1} \Delta^{s}\left(D^{\dagger}\right)^{2s}\left[\Delta^{s} \;{}_{s} u_{-\omega,\ell-m}\right] \label{generalconstr}
 \end{equation}
with $p_s$ is radial Teukolsky-Starobinsky constant for spin $s$ \cite{Casals2021}.
Following the same steps as before and using the coefficients $\mathfrak{C}_s^{(5)}$ and $\mathfrak{C}_s^{(6)}$ in \cite{Casals2021} we find the conditions
 \begin{equation}
 \begin{split}
 {}_{s}\mathcal N^{\ri}_{-\omega,\ell-m}&={}_{-s}\bar{\mathcal N}^{\ri}_{\omega,\ell m}\frac{1}{2s}(2\omega)^{2s}(-1)^{m+s} \\
 {}_{-s}\bar{\mathcal{N}}^{\ru}_{k,\ell m}&=(-1)^{m+s} {}_{s}\mathcal N_{-k,\ell -m}^{\ru} p_s^{-1}\times		\\ &\times\left\{\prod_{j=0}^{2s-1}\left[4kMr_++2i(s-j)\sqrt{M^2-a^2} \right]\right\}.
 \end{split}
 \end{equation} 
 Finally, using the normalization conditions \eqref{innorm} and \eqref{upnorm} we get the result \eqref{normconditions}. The condition \eqref{normconditions} can be fulfilled iff the radial Teukolsky-Starobinsky constant $p_s$ is strictly positive. Fixing $M>0$ and $|a|\leq M$, by Lemma $3.5$ in \cite{Casals2021}, $p_s$ is positive for $s\leq2$. Also by Lemma $3.7$ in the same reference, for $s=2$, the infimum $p_2$ is strictly positive and for $s=1$ it approaches zero for $\omega\to\infty$, $a\neq0$ and a suitable choice of $(\ell,m)$. However, by Lemma $3.10$ in \cite{Casals2021}, for $s=3$, $M>0$ and $0<|a|\leq M$, there is a range of values $(\omega,\ell,m)$ such that $p_3$ is negative. This is expected to hold true for general $s \geq 3$.
 
Hence, for $s=\{1,2\}$ the symplectic normalization is consistent with the constraint \eqref{psiphiconstr}, which is equivalent to the Teukolsky-Starobinsky identity \eqref{eq:RealityCondition}.  For $s=3$, both conditions can not be simultaneously fulfilled, and the same is expected to be the case for all $s \geq 3$.

Also in the CCH approach \cite{Candelas1981}, the NP scalar ${}_{s}\psi$ is quantized, and one can easily check that the mode normalization used there (which, as explained in the Introduction, is determined from the symplectic form of gravitational perturbations or Maxwell fields) coincides with our normalization. 

\subsection{Unruh State} \label{unruhstate}
The limited applications of the Boulware state are well known in literature. For example, it does not describe a black hole formed by gravitational collapse and does not capture Hawking radiation. It is also not Hadamard across the event horizon. It therefore has a diverging expectation value of the renormalized stress energy tensor at the event horizon, limiting its relevance to the extremal limit.

In order to construct the so-called Unruh state in the exterior region of Kerr black hole, we rely (and adapt) the results of \cite{Frolov1989, Casals2005}, where the two-point function in the past Unruh state is given by 
\begin{equation}
\begin{aligned}
	\omega^{\phi\psi}_U(x,x')&:=\langle U|{}_{-s}\phi(x) {}_{s}\psi(x')|U \rangle=\\
	&=\sum_{\ell m}\int_{\mathbb R}\ud k\,  \frac{\mathrm{sgn}(k)}{1-e^{-2\pi k/\kappa}} v^\ru_\lambda(x) u^\ru_{-\lambda}(x')\\& +\SumInt_{\lambda}v^\ri_\lambda(x)u^\ri_{-\lambda}(x')
	\end{aligned}
\end{equation}
where we used the notation introduced below \eqref{state}.  This state corresponds to the absence of particles from $\mathcal I^-$ but the horizon $\mathcal H^-$ is thermally populated, capturing the Hawking radiation phenomenon.

The construction of an Unruh state in the extended Kerr black hole requires additional care due to the fact that our tetrad is singular across the future horizon \cite{Hawking1972, Teukolsky1974}. In particular, it is not evident how to perform an explicit decomposition of Unruh modes into Boulware ones, as performed in e.g. \cite{Klein2021}. We leave the analysis of these issues to a future investigation.
\subsection{Quantization of gravitational and electromagnetic perturbations and zero modes} \label{quantum}

We conclude this section by reconstructing the quantum operators associated to gravitational and electromagnetic fields.
 
In the gravitational case, after the quantization of the Hertz potential ${}_{-2}\phi$, we can reconstruct the linearised metric perturbation associated to it by using \eqref{reco}. It is known that any  perturbation $h_{ab}$ with proper fall-off at infinity (preserving asymptotic flatness), can be expressed (modulo gauge transformation) as
\begin{equation}
h_{ab}=\Re[{}_{2}\mathcal S^\dagger {}_{-2}\phi]_{ab}+\dot{g}_{ab} \label{recometric}
\end{equation}
 where $\dot g_{ab}$ are the zero modes associated to changes of mass or angular momentum, i.e., perturbations towards another Kerr black hole \cite{Wald1973,Green2020}. Thus, we have to discuss the quantization of  $\dot g_{ab}$. One can see that the zero modes are symplectically orthogonal to the metric reconstructed from the Hertz potential w.r.t.\ the symplectic form of linearised gravity \cite{Green2020,Prabhu2018}, and thus can be quantized independently. On the other hand, the zero modes are also symplectically orthogonal among each other (the would-be symplectically dual modes, growing linearly in $t$, do not have the appropriate fall-off at infinity and are thus not in the space of gravitational perturbations to be considered). Hence, we can treat the zero modes as classical perturbations, so in particular, we may set them to $0$, corresponding to fixing (at the linear order) the mass and angular momentum of the perturbation to the ones of the background.

In the electromagnetic case, one should consider the zero mode associated to a change in the charge as well. However, by similar considerations, it is consistent to treat this mode classically and set it to $0$.

To conclude, we can quantize the gravitational and electromagnetic perturbation by only quantizing the Hertz potentials ${}_{-s}\phi$ and reconstructing the field operators using the equation \eqref{reco}.  
 
 \section{Hadamard expansion} \label{pointsplit}
 Once the quantization procedure is understood, we can analyse expectation values of physically meaningful observables. However, as usual in quantum field theory, these expectation values are not well-defined a priori (if the observable is non-linear in the fields). In order to renormalize such quantities, we need to characterize the singular behaviour of the two-point functions, e.g. $\omega_B$ and $\omega_U$. 
 
 In order to perform a Hadamard point-split renormalization, we need to construct the Hadamard parametrices for ${}_{s}\mathcal O$ and ${}_{s}\mathcal O^\dagger$, which encode  the singular behaviours of the two-point functions. At this stage, we emphasize the fact that the TEs \eqref{KGrelation} have the structure of the charged Klein-Gordon equation for a complex external potential $\Gamma^a$. Thus, the Hadamard parametrix for the operator ${}_{s}\mathcal O^\dagger$ can be written as
 \begin{equation}
 H^{\phi\psi}(x,x')=\alpha\left[\frac{U^{\phi\psi}\left(x, x^{\prime}\right)}{\sigma_{\epsilon}\left(x, x^{\prime}\right)}+V^{\phi\psi}\left(x, x^{\prime}\right) \ln \left(-\frac{\sigma_{\epsilon}\left(x, x^{\prime}\right)}{K^{2}}\right)\right] 
 \label{hadparametrix}
 \end{equation}
 where $\alpha=-1/(8\pi^2)$, $K$ is an arbitrary length scale, $\sigma_\epsilon=\sigma-i\epsilon(t-t')$ and $\sigma$ is Synge's world function, equal to half the squared geodesic distance between $x$ and $x'$ \cite{Poisson2011}. The Hadamard coefficients,
 $U(x,x')$ and $V(x,x')$, are smooth functions which are determined in a local and (gauge) covariant way as follows. Writing $V(x,x')=\sum_{n=0}^{+\infty}V_n(x,x')\sigma^n(x,x')$, and requiring that the application of $\mathcal O^\dagger$ yields a smooth function, one derives the transport equations
 	\begin{widetext}
 \begin{align}
& \left[\sigma^a D_a+\frac{1}{2}\Box\sigma-2\right]U^{\phi\psi}  =0, \nonumber\\
\label{syngeexpansion}
 &2\left[\sigma^a D_a+\frac{1}{2} \square \sigma-1\right] V^{\phi\psi}_{0} =-\left[D^a D_a-4s^2\Psi_2\right] U^{\phi\psi}, \nonumber \\
&2(n+1)\left[\sigma^a D_a+\frac{1}{2} \square \sigma+n\right] V^{\phi\psi}_{n+1} =-\left[D^a D_a-4s^2\Psi_2\right] V^{\phi\psi}_{n}, 
 \end{align}
\end{widetext}
 where we introduced $\sigma^a=\nabla^a\sigma$, the covariant derivative $D_a:=(\nabla_a-s\Gamma_a)$ and used $\sigma^a\sigma_a=2\sigma$. Requiring that $U^{\phi\psi}(x,x)=1$ we obtain:
 \begin{equation}
 U^{\phi\psi}(x,x')=\Delta^{\frac{1}{2}}(x,x')P^{\phi\psi}(x,x')
 \end{equation}
 where $\Delta(x,x')$ is the van Vleck-Morette determinant and $P^{\phi\psi}(x,x')$ is the parallel transport with respect to the covariant derivative $D_a$ along the geodesic from $x'$ to $x$. For the applications of the point splitting method, we only need the coinciding point expansion of the Hadamard coefficients. In order to evaluate these, we perform a covariant Taylor expansion of the coefficients in the form
 \begin{equation}
 \begin{aligned}
 K(x,x')=K_{0}(x)&+K_{1a}(x)\sigma^a(x,x')+\\&+K_{2ab}(x)\sigma^a(x,x')\sigma^b(x,x')+\dots
 \end{aligned}
 \end{equation}
 and, adapting the results in \cite{Balakumar2020} for the complex Klein-Gordon field in an external potential, one gets ($s>0$)
 \begin{align}
 U^{\phi\psi}_{0}&=1,\nonumber \\
 U^{\phi\psi}_{1 a}&=s\Gamma_a, \nonumber\\
  U^{\phi\psi}_{2 ab}&=-\frac{s}{2} D_{(a} \Gamma_{b)},\nonumber \\ U^{\phi\psi}_{3 abc}&=\frac{s}{6} D_{(a} D_{b} \Gamma_{c)}, \nonumber\\
 U^{\phi\psi}_{4 abcd}&=\frac{1}{360} R_{(a|f| b}^{e} R_{c|e| d)}^{f}-\frac{s}{24} D_{(a} D_{b} D_{c} \Gamma_{d)}, \label{hadamardexpansionU}
  \end{align}
 and, for the logarithmic part
  \begin{equation}
 \begin{aligned}
 V^{\phi\psi}_{00}=& 2 s^2\Psi_2, \\
 V^{\phi\psi}_{01 a}=&-s^2\Psi_{2; a}+2s^3 \Psi_2 \Gamma_{a} -\frac{1}{12} \nabla^{b} \tilde F_{b a}, \\
 V^{\phi\psi}_{02 ab}=& \frac{s^2}{3} \Psi_{2; ab}-\frac{1}{360} R^{cde}{ }_{a} R_{cdeb} -s^3\Psi_2 D_{(a} \Gamma_{b)}+\\
  &-s^3 \Gamma_{(a} \Psi_{2; b)}+\frac{1}{24} \tilde F^{c}{ }_{a} \tilde F_{b c}+\frac{s}{12} \Gamma_{(a} \nabla^{c} \tilde F_{b) c}+\\
 &-\frac{1}{24} \nabla_{(a} \nabla^{c}\tilde  F_{b) c}, \\
 V^{\phi\psi}_{10}=& 2s^4\Psi_2^{2}-\frac{s^2}{6}\square \Psi_2+ \frac{1}{720} R^{abcd} R_{abcd}+\frac{1}{48}\tilde  F^{ab}\tilde  F_{ab},
 \end{aligned} \label{hadamardexpansionV}
 \end{equation}
 where we defined $\tilde F_{ab}=s(\nabla_a \Gamma_b-\nabla_b \Gamma_a)$. Similarly, the parametrix $H^{\psi\phi}(x,x')$ for ${}_{s}\mathcal O$, i.e., the divergent part of
 \begin{equation}
 	 w^{\psi\phi}(x,x'):=\langle B| \psi(x)\phi(x')|B\rangle=\sum_I\SumInt_\lambda u^I_{\lambda}(x)v^I_{-\lambda}(x'),
 \end{equation}
 can be obtained by simply exchanging $\Gamma_a\to-\Gamma_a$.
 
 In the context of the CCH approach, expectation values of the form $\langle{}_{s}\bar \psi(x){}_{s} \psi(x') \rangle$ were considered for $s=\{\pm1, \pm 2\}$ \cite{ Jensen1995, Casals2005} . Using the Teukolsky-Starobinsky identity \eqref{eq:RealityCondition}, the singular behaviour for $s=\{+1,+2\}$ can easily be obtained as 
 \begin{equation}
 	H^{{}_{s}\bar\psi {}_{s}\psi}(x,x')=\frac{(-i)^{2s}}{2s}\thorn^{2s} H^{\phi\psi}(x,x') \label{npdiv}
 \end{equation}
 and for the opposite spin, one can use the GHP prime transformation defined in Sect. \ref{GHPform} as ${}_{-s}\psi=(i)^{2s}{}_{s}\psi'$ \cite{Geroch1973}, leading to
 \begin{equation}
 		H^{{}_{-s}\bar\psi {}_{-s}\psi}(x,x')=\left(	H^{{}_{s}\bar\psi {}_{s}\psi}(x,x')\right)'. \label{npprimeddiv}
 \end{equation}
In contrast to $H^{\phi\psi}$ and $H^{\psi\phi}$, these are not of Hadamard form: From the derivative in \eqref{npdiv} one obtains a leading divergence in the coinciding point limit of the form $(l^a \sigma_a)^{2s} \sigma^{-2s-1}$  and from \eqref{npprimeddiv} one obtains a leading divergence for $H^{{}_{-s}\bar\psi {}_{-s}\psi}$ of the form $(n^a \sigma_a)^{2s} \sigma^{-2s-1}$. Nevertheless, \eqref{npdiv} and \eqref{npprimeddiv} capture the universal short distance singularity (in Hadamard states) and may thus be used to obtain renormalized expectation values for expressions such as $({}_{\pm s}\bar \psi{}_{\pm s} \psi)(x)$. On the other hand, we note that in our approach it is unclear whether a universal short distance behaviour of $\langle \phi(x)\bar\phi(x') \rangle$ exists. This is a limitation of our framework as it implies that it is presently unclear how to obtain renormalized expectation values for some electromagnetic or gravitational observables. In the electromagnetic case, the only quadratic observable for which this limitation is relevant is $({}_{0}\psi{}_{0}\bar\psi)(x)$ with the electromagnetic NP scalar
${}_{0}\psi=\frac{1}{2}\left(F_{ln}+F_{\bar m m}\right)$. In the gravitational case, quadratic expressions of the Killing invariants $\mathbb I_\xi$ and $\mathbb I_\zeta$ (see \cite{Aksteiner2018}) are affected by this limitation. In any case, for the evaluation of quantum corrections of the gravitational canonical energy, discussed below, only $H^{\phi \psi}$ and $H^{\phi \psi}$ are required for renormalization.
 
 An explicit form for the Hadamard coefficients can be achieved by using standard computer algebra packages. In situations where a locally covariant renormalization scheme is required \cite{Hollands2001}, we will need the coinciding point limit of combinations of $H^{\phi\psi}$, $H^{\psi\phi}$ and their derivatives (e.g. in the renormalization of $\langle{}_{\pm s}\psi {}_{\pm s}\bar \psi \rangle$). In particular, we are only interested in the non-zero contribution coming from the limit $x'\to x$. 
 
  As example of divergent contributions coming, when taking the point split only in time $\Delta x^\alpha=(\tau,\mathbf 0)$, we can expand the Synge's world function in powers of $\tau$ using the formulas in \cite{Ottewill2009}.
  In the coinciding point limit, the Hadamard parametrix for $s=2$ is
 \begin{widetext}
 \begin{equation}
 \begin{split}
 \frac{1}{\alpha}&H^{\phi\psi}=\frac{2 \left(a^2 x^2+r^2\right)}{\tau_\epsilon ^2 \left(a^2 x^2-2 M r+r^2\right)}+\frac{4 \left(a^2 x^2-i a M x-3 M r+r^2\right)}{\tau_\epsilon  (r-i a x) \left(a^2 x^2-2 M r+r^2\right)}+\\+&\frac{M^2 \left\{-a^6 x^4+a^4 r x^2 \left[2 M x^2+r \left(3 x^2-2\right)\right]+a^2 r^3 \left[r \left(2 x^2-1\right)-4 M x^2\right]+r^5 (2 M-r)\right\}}{6 \left(a^2 x^2+r^2\right)^3 \left(a^2 x^2+r (r-2 M)\right)^2}+\\&+\frac{4 \left(a^2 x^2-i a M x-3 M r+r^2\right)^2}{(r-i a x)^3 (r+i a x) \left(a^2 x^2-2 M r+r^2\right)}-\frac{8 M}{\zeta ^3} \log \left[-\frac{\tau^2_\epsilon \left(a^2 x^2-2 M r+r^2\right)}{2 \Sigma }\right]+O\left(\tau ^1\right);
 \end{split}
 \end{equation}
 \end{widetext}
 where  $x=\cos\theta$  and $\tau_{\epsilon}=\tau-i\epsilon$.  In the Schwarzschild limit $a\to0$ this reduces to
 \begin{equation}
 \begin{split}
 \frac{1}{\alpha}H^{\phi\psi}=	&\frac{2 r^2}{\tau ^2_{\varepsilon} \Delta}+\frac{4 (r-3 M)}{ \tau_{\varepsilon} \Delta}+\\+&\frac{215 M^2-144 M r+24 r^2}{6 r^2 \Delta}-\frac{8 M}{ r^3 } \log \left(-\frac{\tau ^2_{\varepsilon} \Delta}{2 r^2}\right)
 \end{split}
 \end{equation}
  with $\Delta= r(r-2 M)$. For the spinless case and $a=0$, we can directly set $s=0$ in the expansions \eqref{hadamardexpansionU} and \eqref{hadamardexpansionV} to get
 \begin{equation}
 \frac{1}{\alpha}H\vert_{s=0}=\frac{2 r}{\tau ^2 _{\varepsilon}(r-2 M)}+\frac{M^2 \left(16 M r^5-8 r^6\right)}{48 r^8 (r-2 M)^2}, \label{schwalimit}
 \end{equation}
 which is the standard result \cite{Levi2015}.

 	\section{Application: Canonical energies} \label{canonicalenergy}
 	\subsection{Classical setting} \label{subsectcanonicalenergy}
 An important non-linear observable in classical linearised gravity on stationary spacetimes is the gravitational canonical energy \cite{Hollands2013}. In this section we show how to use the Teukolsky formalism in order to obtain the divergent part of its expectation value on any Hadamard state. 
 Since we are interested in the gravitational case, we will set $s=2$: we will drop the index $s$ from the previous quantities.
 
  Following \cite{Hollands2013,Prabhu2018}, we construct the gravitational canonical energy as follows. We first implicitly define the vector $w^a$ such that
 \begin{equation}
 h_1^{a b}\left(\mathcal{E}\left[h_2\right]\right)_{a b}-\left(\mathcal{E}^{\dagger}[h_1]\right)_{a b} h_2^{a b}=\nabla_{a} w^{a}\left( h_1, h_2\right),
 \end{equation}
with $\mathcal E$ the equation of motion operator for metric perturbations $h_{ab}$ (recall the discussion in Section~\ref{sec:WaldIdentity}). Then, the symplectic form on the space of solutions (modulo gauge) of the linearised Einstein equation is given by
 \begin{equation}
 \Omega\left(h_1, h_2\right) := \int_{\Sigma} \ud \Sigma_{a} w^{a}\left(h_1, h_2\right),
 \end{equation}
 where $\Sigma$ is a spacelike Cauchy surface.
 Finally, we define the gravitational canonical energy of metric perturbation of Kerr spacetime by setting
 \begin{equation}
 \mathscr E(h):=\Omega(h,\lie{\xi} h)
 \end{equation}
 with $\mathcal L_\xi$ the Lie derivative w.r.t.  $\xi=-(\partial_t)^a-\omega_+(\partial_\varphi)^a$.
 
For the canonical energy to be physically meaningful -- and for it to be gauge invariant --
certain gauge conditions must be chosen at the horizon bifurcation cross section and certain asymptotic conditions must hold near spatial infinity
\cite{Hollands2013}. Under these conditions, the canonical energy is related to second order corrections to the black hole parameters \cite{Hollands2013} 
by the master formula
\begin{equation}
\mathscr{E}(h)=\delta^{2} M- \omega_+ \delta^{2} J-\frac{\kappa}{8 \pi} \delta^{2} A. \label{classicalcanonical}
\end{equation}
The detailed properties of the canonical energy were not investigated in \cite{Hollands2013}  in the case of extremal black holes. For the sake of our informal discussion below, we shall assume that the conditions hold in the extremal limit by some sort of continuity; in particular using 
$\kappa = 0,\, \omega_+=1/(2M),\, M^4=J^2$ in the extremal case, the 
master formula would become 
\begin{equation}
\mathscr{E}(h)=\frac{\delta^{2} (M^4-J^2)}{4M^3}
\label{classicalcanonical2}
\end{equation}
for any perturbation satisfying $\delta M = \delta J = 0$. From \eqref{classicalcanonical2} it is evident that a negative sign of the canonical energy implies instability of the black hole, i.e., the black hole is overspinning.

In \cite{Hollands2013}, an expression for the canonical energy was given directly in terms of the perturbation $h_{ab}$ which however 
	is rather complicated.
	
On the other hand, we can also use the symplectic form \eqref{symp} to define the canonical energy of Teukolsky fields as
 \begin{equation}
 \mathscr E_T(\Psi)=\Re\tilde\sigma(\Psi,\GHPLie_{\xi}\Psi). \label{canonicallie}
 \end{equation} 
 Here $\GHPLie_{\xi}$ is the GHP generalization of $\mathcal L_\xi$ and it reduces to the standard Lie derivative when applied on GHP quantity of type $\GHPw{0}{0}$ and for $\eta\GHPwt\GHPw{p}{q}$. Its explicit definition is given in App. \ref{GHPlie}.
 
 Assume that $h=\Re \mathcal S^\dagger \phi$ and thus $\Psi=(\phi,\mathcal T \Re (\mathcal S^\dagger \phi))$. We want to understand the relation between
 \begin{equation}
 \mathscr E(\Re \mathcal S^\dagger \phi) \quad \text{ and }\quad \mathscr E_T(\Psi).
 \end{equation}
For this, it is advantageous to express the symplectic form $\tilde \sigma$ as
 \begin{equation}
 \tilde\sigma(\Psi_1,\Psi_2)=\Pi(\phi_1,\psi_2)-\Pi(\phi_2,\psi_1),
 \end{equation}
with 
\begin{align}
 \Pi(\phi,\psi) & :=\int_\Sigma u_a\pi^a(\phi,\psi), \\
\label{pia}
 \pi^{a}\left(\phi, \psi\right) & := \phi(\Theta^a+4B^a)\psi-\psi(\Theta^a-4B^a)\phi.
\end{align} 
 If $\phi$ is a smooth solution to $\mathcal O^\dagger\phi=0$ with initial data of compact support on some Cauchy surface and if $h$ is a smooth, real perturbation solving the linearised Einstein equation $\mathcal E h=0$, then \cite{Prabhu2018} 
 	\begin{equation}
 	\Omega(\Re(\mathcal S^\dagger \phi),h)=\Re(\Pi(\phi,\mathcal T h)). \label{propwald}
 \end{equation}
 This intimate relation between the symplectic forms for Teukolsky fields and metric perturbations explains the fact that mode normalization in the CCH approach coincides with the normalization in Sect. \ref{quantization}.

Using that in the Kinnersley tetrad $\GHPLie_{\xi}$ annihilates all the legs and spin coefficients and commutes with all GHP operators \cite{Collinson1990} and \eqref{propwald}, we finally obtain
 \begin{equation}
 \begin{split}
 \mathscr E_T(\Psi)&=\Re \tilde\sigma(\Psi,\GHPLie_{\xi}\Psi)\\&=\Re\left[\Pi(\phi,\GHPLie_{\xi}\psi)-\Pi(\GHPLie_{\xi}\phi,\psi)\right]\\
 &=\Omega(\Re\mathcal S^\dagger \phi,\mathcal L_\xi \Re \mathcal S^\dagger\phi)-\Omega(\Re\mathcal L_\xi \mathcal S^\dagger \phi,\Re S^\dagger \phi)\\&=
 2\mathscr E(\Re (\mathcal S^\dagger\phi))=2 \mathscr E(h)
 \end{split}\label{eq:E_H=2E}
 \end{equation} 
 when $\psi=\mathcal T\Re (\mathcal S^\dagger \phi)$, i.e. when the Teukolsky fields $\Psi = (\psi, \phi)$ fulfill the constraint \eqref{psiphiconstr} and $h=\Re \left(\mathcal{S}^\dagger\phi\right)$. In the above derivation, we assumed that $\phi$ is of compact support on the given Cauchy surface.
 
 \subsection{Canonical energy operator}
In the context of quantum field theory on Kerr spacetime, it is tempting to replace the classical expression \eqref{classicalcanonical} and \eqref{classicalcanonical2} by an expectation value $\langle \Phi|\mathscr E(h)|\Phi \rangle$ in order to understand how vacuum fluctuations would affect the balance between mass, angular momentum and area. 

It is already known that classical gravitational fluctuations with $\delta M=\delta J=0$ cannot achieve a negative sign on the right side of \eqref{classicalcanonical} or \eqref{classicalcanonical2} \cite{Wald78,Sorce2017}. However, this leaves open the possibility of a negative sign of $\langle \Phi|\mathscr E(h)|\Phi \rangle$ due to quantum effects. Thus, a first principle calculation seems necessary.

 Unfortunately, the computation of the canonical energy $\langle \Phi| \mathscr E(h) | \Phi \rangle$, even for the Boulware state, is a daunting task: Not only does it have a complicated algebraic structure \cite{Hollands2013}, but even worse, when the metric perturbation $h$ is reconstructed from the Hertz potential $\phi$, one obtains a bilinear in $\phi$ involving up to six derivatives (one from the symplectic form, one from the time derivative and two times two from the reconstruction). To renormalize such an energy density, one would have to go to very high order in the Hadamard expansion. Hence, the canonical energy \eqref{classicalcanonical} in the form given in \cite{Hollands2013} does not seem to be a useful starting point for the evaluation in the quantum field theory.
 
Instead, one may attempt to start with the expression for the canonical energy in terms of the Teukolsky fields given in the above subsection \ref{subsectcanonicalenergy}. For a Teukolsky field satisfying the constraint \eqref{eq:RealityCondition} and expanded in symplectically normalized modes as in \eqref{expansion}, one obtains
\begin{widetext}
 \begin{equation}
 \begin{split}
 \tilde \sigma(\Psi,\GHPLie_{\xi}\Psi)=\sum_{\ell, m}&\left(
 \int_{0}^{\infty}\ud\omega\, k(a^{\ri\dagger}_{\omega,\ell m}a^{\ri}_{\omega,\ell m}+a^{\ri}_{\omega,\ell m}a^{\ri\dagger}_{\omega,\ell m}+b^{\ri\dagger}_{\omega,\ell m}b^{\ri}_{\omega,\ell m}+b^{\ri}_{\omega,\ell m}b^{\ri\dagger}_{\omega,\ell m})\right.\\
 &\left.+
 \int_{0}^{\infty}\ud k\, k(a^{\ru\dagger}_{k,\ell m}a^{\ru}_{k,\ell m}+a^{\ru}_{k,\ell m}a^{\ru \dagger}_{k,\ell m}+b^{\ru\dagger}_{k,\ell m}b^{\ru}_{k,\ell m}+b^{\ru}_{k,\ell m}b^{\ru\dagger}_{\omega,\ell m})
 \right).
 \end{split} \label{energyinnerprod}
 \end{equation}
\end{widetext} 
This can be shown by evaluating the symplectic form on $\mathcal I^-$ and $\mathcal H^-$, so that in particular the $\ri$- and the $\ru$-modes are manifestly orthogonal. One also uses \eqref{eq:ModeConstraint} and the fact that in the Kinnersley frame $\GHPLie_{\xi}=-\partial_t-\omega_+\partial_{\varphi}$ on GHP scalars \cite{Pound2021}, so that
 \begin{equation}
 \GHPLie_{\xi} {}_{2}u^{\ri / \ru}_{\omega, \ell m}=ik \ {}_{2}u^{\ri / \ru}_{\omega, \ell m}.
 \end{equation}

\noindent Straightforward computations, will then lead to \eqref{energyinnerprod}.
 
 According to the principles of QFTCS \cite{WaldQFTCS, Hollands2015}, the renormalization of non-linear observables must be performed locally, i.e., we need to renormalize the associated density. From the definition of $\mathscr E_T$, we can read off the canonical energy density directly from the integral
 \begin{equation}
 \begin{split}
 \mathscr E_T(\Psi)&=\Re\left(\Pi(\phi,\GHPLie_{\xi}\psi)-\Pi(\GHPLie_{\xi}\phi,\psi)\right)\\
 &=\int_\Sigma \Re \left( u_a\pi^a(\phi,\GHPLie_{\xi}\psi)-u_a\pi^a(\GHPLie_{\xi}\phi,\psi) \right) \ud \Sigma \label{totalenergy}
 \end{split}
 \end{equation}
 where $u_a$ is given by \eqref{normal}. In terms of modes, we compute the density
 \begin{equation}
 \begin{split}
 &\langle B|u^a\pi_a(\phi,\GHPLie_{\xi}\psi)|B\rangle=\\
 &=\sum_I\SumInt_{\lambda}\bigl[
 k(u^t\omega-u^\varphi m-2iu_a\Gamma^a ){}_{-2}v^I_\lambda{}_{2}u^I_{-\lambda}+\bigr.\\\bigl.&\hspace{1cm}+k(u^t\omega-u^\varphi m+2iu_a\Gamma^a ){}_{2}u^I_\lambda{}_{-2}v^I_{-\lambda}	\bigr]
 \end{split}
 \end{equation}
 where \eqref{paststate} have been used.
 Finally, it easy to check also that
 \begin{equation}
 \langle u^a\pi_a(\phi,\GHPLie_{\xi}\psi)\rangle=-\langle u^a\pi_a(\GHPLie_{\xi}\phi,\psi)\rangle
 \end{equation}
  and thus the divergent expectation value of the canonical energy density is given by
 \begin{equation}
 \begin{split}
 &\Re	\langle B| u^a[\pi_a(\phi,\GHPLie_{\xi}\psi)-\pi_a(\GHPLie_{\xi}\phi,\psi)]|B\rangle=\\&=2\Re \sum_I\SumInt_{\lambda}k\bigl[
 (u^t\omega-u^\varphi m-2iu_a\Gamma^a ){}_{-2}v^I_\lambda{}_{2}u^I_{-\lambda}+\bigr.\\\bigl.&\hspace{1cm}+(u^t\omega-u^\varphi m+2iu_a\Gamma^a ){}_{2}u^I_\lambda{}_{-2}v^I_{-\lambda}	\bigr]. \label{density}
 \end{split}
 \end{equation}
 In a completely analogous manner, one can show
 \begin{equation}
 	\begin{split}
 	&\langle U | u^a[\pi_a(\phi,\GHPLie_{\xi}\psi)-\pi_a(\GHPLie_{\xi}\phi,\psi)] |U \rangle\\
 	&=2\Re\SumInt_{\lambda}k\bigl[
 	(u^t\omega-u^\varphi m-2iu_a\Gamma^a ){}_{-2}v^\ri_\lambda{}_{2}u^\ri_{-\lambda}+\bigr.\\\bigl.&\hspace{1cm}+(u^t\omega-u^\varphi m+2iu_a\Gamma^a ){}_{2}u^\ri_\lambda{}_{-2}v^\ri_{-\lambda}	\bigr]+\\&+2\Re\SumInt_{\lambda}	k\coth\left(\frac{\pi k}{\kappa}\right)\bigl[
 (u^t\omega-u^\varphi m-2iu_a\Gamma^a ){}_{-2}v^\ru_\lambda{}_{2}u^\ru_{-\lambda}+\bigr.\\\bigl.&\hspace{1cm}+(u^t\omega-u^\varphi m+2iu_a\Gamma^a ){}_{2}u^\ru_\lambda{}_{-2}v^\ru_{-\lambda}	\bigr].
 	\end{split}
 \end{equation}
 The renormalization of these quantities proceeds with standard techniques of QFTCS. The canonical energy density $ e_T(x)$ for the Teukolsky fields can be read from the integrand of \eqref{totalenergy} and the definition of $\pi_a$ \eqref{pia}, namely
 \begin{equation}
 \label{eq:e_H}
 \begin{split}
  e_T(x)=-u^a\Re&[\phi(x)D^\dagger_a \xi^b\partial_b \psi(x)-\xi^b\partial_b \psi(x) D_a \phi(x)\\&-\xi^b\partial_b \phi(x) D^\dagger_a\psi(x)+\psi(x)D_a \xi^b\partial_b \phi(x)],
 \end{split}
 \end{equation}
 with $D^\dagger_a=\nabla_a+2\Gamma_a$. After quantization, i.e., understanding $\phi$ and $\psi$ as quantum fields, the previous expression is no longer well defined since it contains pointwise products between fields and their derivatives. In the point-split method, we start from\footnote{In principle, this expression is ill-behaved under gauge transformations. In order to ensure gauge invariance, one can introduce the parallel transports $P^{\phi\psi}(x,x')$ and $P^{\psi\phi}(x,x')$ in the expressions \eqref{pointsplitenergy} and \eqref{hadamardenergy}. However, in view of the limit \eqref{coinclimitenergy}, the introduction of parallel transports will not affect the coincidence point limit (see discussion in \cite{Wernersson2021}).}
 \begin{align}
\label{pointsplitenergy}
 & \langle e_T(x,x')\rangle \\
 & = - u^a\xi^b \Re \langle \left[ D^\dagger_{a} \partial_{b} \psi(x)\phi(x') + D_{a}\partial_{b}\phi(x) \psi(x') \right. \nonumber \\
 & \ \ \left. -\partial_{b} \psi(x) g_{\alpha}{}^{\alpha'} D_{a'}\phi(x')  -  \partial_{b}\phi(x) g_{\alpha}{}^{\alpha'} D^\dagger_{a'}\psi(x')\nonumber \right] \rangle.
 \end{align} 
with $ g_{\alpha}{}^{\alpha'}=g_{\alpha}{}^{\alpha'}(x,x')$ the parallel transport of vectors from $x'$ to $x$ \cite{Poisson2011}. 
The universal short distance singularity of the above can be obtained by evaluating
\begin{align}
 -u^a \xi^b& \left[ D^\dagger_a \partial_b H^{\psi\phi}(x,x') + D_a \partial_b H^{\phi\psi}(x,x') \right.\nonumber \\
 & \left. -  g_{\alpha}{}^{\alpha'}D_{a'} \partial_{b} H^{\psi\phi}(x,x') - g_{\alpha}{}^{\alpha'} D^\dagger_{a'} \partial_{b} H^{\phi\psi}(x,x') \right].  \label{hadamardenergy}
\end{align}
We give an explicit expansion of the divergent part $\mathcal D$ in App.~\ref{HadamardExp} (for a point-split in time direction). Finally, the regularized expectation value can be obtained via
 \begin{equation}
 	\langle e_T (x)\rangle_{\mathrm{ren}}:=\lim\limits_{x'\to x}\left[\langle e_T(x,x')\rangle-\Re\mathcal D(x,x')\right]. \label{coinclimitenergy}
 \end{equation}
The short distance singularity to be renormalized here has exactly the same form as the $u^a \xi^b T_{ab}$ component of the stress tensor of the Klein-Gordon field. Hence, a numerical implementation should be possible in situations for which the latter can be handled. This applies to Schwarzschild spacetime \cite{Levi2015,Levi2017}, but, to the best of knowledge, not yet to Kerr spacetime (but note that a stress tensor renormalization has been performed in the interior region \cite{Zilberman2022}). We do not attempt an evaluation here, but conclude this section with few remarks:

\begin{itemize}
\item In order to ensure the conservation of the canonical energy at the quantum level, one has to make sure that the renormalized current, i.e. \eqref{eq:e_H} without the contraction with $u^a$, is also conserved. In particular, by using the facts that the coefficients \eqref{hadamardexpansionV} are $t$- and $\varphi-$independent and that $V_{10}^{\phi\psi}=V_{10}^{\psi\phi}$, from formulas in \cite{Zahn2015} it follows
\begin{equation}
	\nabla^a \langle e_{a} \rangle_{\mathrm{ren}}=0.
\end{equation}
\item As it is obvious from \eqref{energyinnerprod}, the Boulware state is not a stable state. Because of superradiance, there exist in-modes with $k<0$ even for positive $\omega$, making the Teukolsky canonical energy unbounded from below. In order consider a stable state, one has to construct it with $k>0$ also in $\mathcal I^-$. States with similar properties have been constructed in \cite{Balakumar2022} in the Reissner-Nordstrom black hole. One can adapt that construction to our case in order to obtain a meaningful state to analyse the quantum stability of a Kerr black hole.
\item The identification \eqref{eq:E_H=2E} of the canonical energies of Teukolsky fields and metric perturbations was only proven for fields with compactly supported Cauchy data. Quantum fluctuations are not restricted in this way, so in order to compute the gravitational canonical energy by expressing it via the ``Teukolsky canonical energy'', also boundary terms need to be taken into account, which can be obtained from the boundary terms in the relation between the symplectic forms for metric perturbations and Teukolsky fields \cite{Green2022}.
\end{itemize}

  \section{Concluding Remarks}
 
We discussed the canonical quantization of gravitational and electromagnetic perturbations on Kerr spacetime. We showed how to construct field operators in a fixed gauge and that quantum states can be obtained as for a Klein-Gordon theory in an external potential. Our construction is based on Teukolsky fields, i.e., of pairs $\Psi = (\phi, \psi)$ of a Hertz potential $\phi$ and an NP scalar $\psi$. We showed that the (Teukolsky-Starobinsky) constraint relating $\phi$ and $\psi$ can be implemented for spin $s \leq 2$, i.e., for the Hertz potentials corresponding to metric perturbations and Maxwell fields. However, for higher spin the constraint can not be implemented (for $s=3$, this follows from results of \cite{Casals2021}, but it is expected that these extend to all $s \geq 3$).

Our quantization of the Hertz potential is equivalent to that obtained in the CCH approach \cite{Candelas1981}, but we think that our approach has conceptual as well as practical advantages: On the conceptual side, we do not have to split the metric perturbation (or the vector potential) into two parts which are reconstructed from the Hertz potential in different gauges. Instead, we can reconstruct the metric perturbation in a well-defined gauge. On the practical side, we have not only directly quantized a relevant observable, an NP scalar, but also obtained a Hadamard parametrix, which can be used to perform a Hadamard point-split renormalization of physically relevant quantities. As a further potential application of our approach, we worked out the relation between the canonical energy of gravitational perturbations and that of Teukolsky fields (up to boundary terms), which may allow for the computation of semiclassical corrections to black hole masses (and thus address the issue of quantum stability of extremal black holes).
 
In a future work, we aim to apply the results developed here to an explicit computation of propagators and expectation values of local and gauge invariant observables in the electromagnetic/gravitational perturbations in order to have an explicit gauge invariant characterization of such fluctuations around Kerr spacetime. Furthermore, we aim to make progress on the computation of the canonical energy of gravitational perturbations.
 
 \begin{acknowledgments}
 	C.I is grateful to the International Max Planck Research School for Mathematics in the Sciences, Leipzig for supporting this work. We would like to thank S. Hollands for useful discussions and comments. This work makes use of the Black Hole Perturbation Toolkit.
 \end{acknowledgments}
 \appendix
 	\section{Kinnersley frame} \label{kinnframe}
 In Boyer-Lindquist coordinates, the Kinnersley frame reads \cite{Kinnersley,Teukolsky1973}:
 \begin{equation}
 \begin{aligned}
 l^{a} &=\frac{1}{\Delta}\left(r^{2}+a^{2}, \Delta,0, a\right), \\ n^{a}&=\frac{1}{2 \Sigma}\left(r^{2}+a^{2},-\Delta, 0, a\right) \\
 m^{a} &=\frac{1}{2^{1 / 2}(r+i a \cos \theta)}\left(i a \sin \theta, 0,1, i / \sin \theta\right)\label{kinnersley}
 \end{aligned}
 \end{equation}
 and the non-vanishing spin-coefficients:
 \begin{align}
 &\rho=-\frac{1}{\zeta},&  &\rho^{\prime}=\frac{\Delta}{2 \zeta^{2} \bar{\zeta}}, \\ &\tau=-\frac{i a \sin \theta}{\sqrt{2} \zeta \bar{\zeta}},& &\tau^{\prime}=-\frac{i a \sin \theta}{\sqrt{2} \zeta^{2}} \\
 &\beta=\frac{\cot \theta}{2 \sqrt{2} \bar{\zeta}}, & &\beta^{\prime}=\frac{\cot \theta}{2 \sqrt{2} \zeta}-\frac{i a \sin \theta}{\sqrt{2} \zeta^{2}}, \\ &\varepsilon=0, & &\varepsilon^{\prime}=\frac{\Delta}{2 \zeta^{2} \bar{\zeta}}-\frac{r-M}{2 \zeta \bar{\zeta}}\label{spincoeff}
 \end{align}
 with $\zeta=r-ia\cos\theta$. 
 
 \section{The GHP Lie derivative $\GHPLie_{\xi}$}  \label{GHPlie}
  The Lie derivative $\GHPLie_{\xi}$ in \eqref{canonicallie} can be defined on GHP tensors as
 \begin{equation}
 \GHPLie_{\xi}:=\GHPLie_{t}+\omega_+\GHPLie_{\varphi}
 \end{equation}
 with \cite{Pound2021}
 \begin{equation}
 \begin{split}
 &\GHPLie_{t}=\mathcal L^\Theta_t+\frac{p}{2}\zeta\Psi_2+\frac{q}{2}\bar\zeta\bar{\Psi}_2; \\
 &\GHPLie_{\varphi}=\frac{1}{a}\GHPLie_{\Xi}-a \GHPLie_{t};\\
 &\GHPLie_{\Xi}=\frac{\zeta}{4}\left[(\zeta-\bar{\zeta})^2\left(\rho^{\prime} \thorn-\rho \thorn^{\prime}\right)-(\zeta+\bar{\zeta})^2\left(\tau^{\prime} \edth^{\prime}-\tau \edth\right)\right]\\&\hspace{1cm} -p\, {}_\Xi h_1-q\,{}_\Xi \bar{h}_1;\\
 &{ }_\Xi h_1=\frac{1}{8} \zeta\left(\zeta^2+\bar{\zeta}^2\right) \Psi_2-\frac{1}{4} \zeta \bar{\zeta}^2 \bar{\Psi}_2+\\&\hspace{1.5cm}+\frac{1}{2} \rho \rho^{\prime} \zeta^2(\bar{\zeta}-\zeta)+\frac{1}{2} \tau \tau^{\prime} \zeta^2(\bar{\zeta}+\zeta).
 \end{split}
 \end{equation}
 The action of $\mathcal L^\Theta_t$ on a tensor field  $\eta_{b_1 \ldots b_l}^{a_1 \ldots a_k}\GHPwt\GHPw{p}{q}$ is
 \begin{eqnarray}
 \mathcal L^\Theta_t \eta_{b_1 \ldots b_l}^{a_1 \ldots a_k}&=&t^c \Theta_c \eta_{b_1 \ldots b_l}^{a_1 \ldots a_k}-\sum_{j=1}^k \Theta_c t^{a_j} \eta_{b_1 \ldots b_l}^{a_1 \ldots a_k}+\nonumber\\ & &+\sum_{j=1}^l \Theta_{b_j} t^c \eta_{b_1 \ldots c \ldots b_l}^{a_1 \ldots a_k}
 \end{eqnarray}
 where we may express the generator of time translations as
 \begin{equation}
 t^a=\zeta(-\rho' l^a+\rho n^a+\tau'm^a-\tau\bar m^a).
 \end{equation}
 Observe that also the operator $\GHPLie_{\Xi}$ can be associated to a generator of symmetries in Kerr \cite{Carter1968} (see also \cite{Pound2021}).
 
 \section{Divergences of the canonical energy density}\label{HadamardExp}
 
  Here we give an explicit expression of the divergent part of:
  
 \begin{equation}
 \begin{split}
 \mathcal D:=-u^{a}\nabla_{a}\partial_{t}H_S(x,x')+&\partial_t u^{a'}\nabla_{a'}H_S(x,x')+\\&+ 4u_t \Gamma^t \partial_t H_A(x,x')
 \end{split} \label{divergentD}
 \end{equation}
where $H_S=H^{\phi\psi}+H^{\psi\phi}$,  $H_A=H^{\phi\psi}-H^{\psi\phi}$. By similar and more tedious computations, one can achieve the divergent part related to the angular derivative $\xi^\varphi \partial_\varphi$.  In particular, separating points in the $t-$direction, we can write the non-vanishing part in the limit $\tau\to0$ as:
\begin{equation}
\begin{split}
\mathcal D_{\text{div}}=\frac{1}{8\pi^2}&\Biggl[\left(\frac{1}{\tau_\epsilon^4}\right)A(r,\theta)+\left(\frac{1}{\tau^2_\epsilon}\right)B(r,\theta)+\Biggr.\\&\Biggl.+\ln(-f(r,\theta)\tau_\epsilon^2)C(r,\theta)+D(r,\theta)\Biggr]
	\end{split}
\end{equation}
 $A,B,C,D$ are smooth functions determined by the expansion \eqref{syngeexpansion},\eqref{hadamardexpansionU}, \eqref{hadamardexpansionV} and their derivatives and $f(r,\theta)=\left(a^2 x^2-2 M r+r^2\right)/[2 \left(a^2 x^2+r^2\right)]$, with $x=\cos\theta$. In particular, using the Black Hole Perturbation Toolkit for Mathematica \cite{BHPToolkit}:
\begin{widetext}
	\begin{equation}
 	\begin{split}
 	A:&\qquad \frac{48 \Sigma ^2}{(\Sigma -2 M r)^2} \sqrt{\frac{\Delta  \Sigma }{\Gamma }};\\
 	B:&\qquad \frac{16}{3\zeta\Sigma (\Sigma -2 M r)^3} \sqrt{\frac{\Delta  \Sigma }{\Gamma }} \left(6 i a^7 x^7+6 a^6 r x^6-6 i a^5 x^5 \left(M^2+6 M r-3 r^2\right)+18 a^4 r x^4 \left(-M^2-2 M r+r^2\right)+\right.\\&\left.\hspace{-1.5cm}+i a^3 r x \left(M^3 \left(11 x^2+1\right)+84 M^2 r x^2-72 M r^2 x^2+18 r^3 x^2\right)+a^2 r^2 \left(M^3 \left(37 x^2-1\right)+60 M^2 r x^2-72 M r^2 x^2+18 r^3 x^2\right)+\right.\\&\left.+6 i a r^3 x \left(-14 M^3+15 M^2 r-6 M r^2+r^3\right)+6 r^4 \left(-10 M^3+13 M^2 r-6 M r^2+r^3\right)\right);\\
 	C:&\qquad \frac{16 M}{\zeta^3 \Sigma^3} \sqrt{\frac{\Delta  \Sigma }{\Gamma }} \left(2 a^4 x^4-i a^3 \left(M \left(9 x^2+2\right) x+4 r x^3\right)+a^2 M r \left(2-35 x^2\right)+i a r^2 x (37 M-4 r)+r^3 (15 M-2 r)\right),
 	\end{split}
 	\end{equation}
 	and finally:
 	\begin{equation}
 		\begin{split}
 	D:&\qquad -\frac{1}{45 \Sigma ^6 (\Sigma -2 M r)^4}\left(480 a^{16} x^{16}+120 a^{15} i \left(71 M x^2-16 r x^2+M\right) x^{13}-240 i a^{13} \left(\left(16 x^2+3\right) M^3+\right.\right.\\&\left.\left.\hspace{-1.5cm}+r \left(1105 x^2+46\right) M^2+3 r^2 \left(28 x^2-1\right) M+40 r^3 x^2\right) x^{11}+12 a^{14} M \left(10 r \left(309 x^2+1\right) x^2+M \left(2370 x^4+413 x^2-3\right)\right) x^{10}+\right.\\&\left.\hspace{-1.5cm}+120 a^{11} i r \left(4 \left(37 x^2+12\right) M^4+2 r \left(7735 x^2+333\right) M^3+4 r^2 \left(623 x^2-115\right) M^2+r^3 \left(15-1739 x^2\right) M-144 r^4 x^2\right) x^9+\right.\\&\left.\hspace{-1.5cm}+a^{12} \left(5 \left(96 x^4-1\right) M^4-48 r \left(5340 x^4+841 x^2-6\right) M^3-12 r^2 \left(62166 x^4-1552 x^2+15\right) M^2+240 r^3 x^2 \left(559 x^2+3\right) M+\right.\right.\\&\left.\left.\hspace{-1.5cm}-9600 r^4 x^4\right) x^8-240 i a^9 r^2 \left(4 \left(28 x^2+13\right) M^5+34 r \left(669 x^2+28\right) M^4+r^2 \left(10873 x^2-1362\right) M^3+r^3 \left(460-13933 x^2\right) M^2+\right.\right.\\&\left.\left.\hspace{-1.5cm}+5 r^4 \left(369 x^2-2\right) M+40 r^5 x^2\right) x^7-2 a^{10} r \left(10 \left(72 x^4+25 x^2-2\right) M^5+r \left(-419352 x^4-63663 x^2+370\right) M^4+\right.\right.\\&\left.\left.\hspace{-1.5cm}-24 r^2 \left(97747 x^4-2453 x^2+24\right) M^3+30 r^3 \left(35301 x^4-313 x^2+6\right) M^2-60 r^4 x^2 \left(979 x^2+15\right) M+15360 r^5 x^4\right) x^6+\right.\\&\left.\hspace{-1.5cm}+120 a^7 i r^3 \left(16 \left(7 x^2+4\right) M^6+8 r \left(7647 x^2+319\right) M^5+24 r^2 \left(3841 x^2-250\right) M^4+12 r^3 \left(343-11143 x^2\right) M^3+\right.\right.\\&\left.\left.\hspace{-1.5cm}+8 r^4 \left(5293 x^2-115\right) M^2-5 r^5 \left(643 x^2-3\right) M+80 r^6 x^2\right) x^5-a^8 r^2 \left(-20 x^2 \left(51 x^2+44\right) M^6+12 r \left(99083 x^4+15140 x^2+\right.\right.\right.\\&\left.\left.\left.\hspace{-1.5cm}-58\right) M^5+r^2 \left(13505661 x^4-256304 x^2+2190\right) M^4-48 r^3 \left(184093 x^4-1402 x^2+36\right) M^3+12 r^4 \left(44541 x^4+1000 x^2+\right.\right.\right.\\&\left.\left.\left.\hspace{-1.5cm}+30\right) M^2+2400 r^5 x^2 \left(53 x^2-1\right) M+43200 r^6 x^4\right) x^4-240 i a^5 r^5 \left(8 \left(1943 x^2+84\right) M^6+4 r \left(21321 x^2-677\right) M^5+\right.\right.\\&\left.\left.\hspace{-1.5cm}-4 r^2 \left(32153 x^2-762\right) M^4+r^3 \left(57746 x^2-1377\right) M^3+r^4 \left(230-8953 x^2\right) M^2+r^5 \left(442 x^2-3\right) M-72 r^6 x^2\right) x^3+\right.\\&\left.\hspace{-1.5cm}+4 a^6 r^4 \left(4 x^2 \left(38736 x^2+6029\right) M^6+2 r \left(2307539 x^4-24621 x^2+159\right) M^5+r^2 \left(-2923083 x^4+1239 x^2-545\right) M^4+\right.\right.\\&\left.\left.\hspace{-1.5cm}+24 r^3 \left(-38173 x^4+1051 x^2+12\right) M^3+3 r^4 \left(260616 x^4-3065 x^2-15\right) M^2-150 r^5 x^2 \left(497 x^2-3\right) M-7680 r^6 x^4\right) x^2+\right.\\&\left.\hspace{-1.5cm}+240 a i r^9 (r-2 M)^2 \left(-4698 M^4+5179 r M^3-1673 r^2 M^2+115 r^3 M+8 r^4\right) x+120 a^3 i (2 M-r) r^7 \left(8 \left(7045 x^2-88\right) M^5+\right.\right.\\&\left.\left.\hspace{-1.5cm}+4 r \left(257-12074 x^2\right) M^4+2 r^2 \left(1723 x^2-255\right) M^3+6 r^3 \left(757 x^2+15\right) M^2-r^4 \left(467 x^2+1\right) M-80 r^5 x^2\right) x+\right.\\&\left.\hspace{-1.5cm}+r^{10} (r-2 M)^2 \left(-203573 M^4+226368 r M^3-73692 r^2 M^2+4800 r^3 M+480 r^4\right)+\right.\\&\left.\hspace{-1.5cm}+2 a^2 M (2 M-r) r^8 \left(4 \left(664618 x^2-5141\right) M^4+r \left(30941-3782242 x^2\right) M^3+36 r^2 \left(51813 x^2-443\right) M^2+\right.\right.\\&\left.\left.\hspace{-1.5cm}-6 r^3 \left(59395 x^2-493\right) M+60 r^4 \left(307 x^2-1\right)\right)+a^4 r^6 \left(\left(13328 x^2-9716936 x^4\right) M^6-8 r \left(314726 x^4-18638 x^2-77\right) M^5+\right.\right.\\&\left.\left.\hspace{-1.5cm}+r^2 \left(19080922 x^4-249696 x^2-725\right) M^4-48 r^3 \left(276026 x^4-2803 x^2-6\right) M^3+12 r^4 \left(266304 x^4-2152 x^2-3\right) M^2+\right.\right.\\&\left.\left.\hspace{-1.5cm}-240 r^5 x^2 \left(793 x^2-3\right) M-9600 r^6 x^4\right)\right) \sqrt{\frac{\Delta  \Sigma }{\Gamma }}.
 	\end{split}
 	\end{equation}
 \end{widetext}

\bibliography{MassCorrectionKerr,BHPToolkit}

\begin{thebibliography}{50}%
\makeatletter
\providecommand \@ifxundefined [1]{%
 \@ifx{#1\undefined}
}%
\providecommand \@ifnum [1]{%
 \ifnum #1\expandafter \@firstoftwo
 \else \expandafter \@secondoftwo
 \fi
}%
\providecommand \@ifx [1]{%
 \ifx #1\expandafter \@firstoftwo
 \else \expandafter \@secondoftwo
 \fi
}%
\providecommand \natexlab [1]{#1}%
\providecommand \enquote  [1]{``#1''}%
\providecommand \bibnamefont  [1]{#1}%
\providecommand \bibfnamefont [1]{#1}%
\providecommand \citenamefont [1]{#1}%
\providecommand \href@noop [0]{\@secondoftwo}%
\providecommand \href [0]{\begingroup \@sanitize@url \@href}%
\providecommand \@href[1]{\@@startlink{#1}\@@href}%
\providecommand \@@href[1]{\endgroup#1\@@endlink}%
\providecommand \@sanitize@url [0]{\catcode `\\12\catcode `\$12\catcode
  `\&12\catcode `\#12\catcode `\^12\catcode `\_12\catcode `\%12\relax}%
\providecommand \@@startlink[1]{}%
\providecommand \@@endlink[0]{}%
\providecommand \url  [0]{\begingroup\@sanitize@url \@url }%
\providecommand \@url [1]{\endgroup\@href {#1}{\urlprefix }}%
\providecommand \urlprefix  [0]{URL }%
\providecommand \Eprint [0]{\href }%
\providecommand \doibase [0]{https://doi.org/}%
\providecommand \selectlanguage [0]{\@gobble}%
\providecommand \bibinfo  [0]{\@secondoftwo}%
\providecommand \bibfield  [0]{\@secondoftwo}%
\providecommand \translation [1]{[#1]}%
\providecommand \BibitemOpen [0]{}%
\providecommand \bibitemStop [0]{}%
\providecommand \bibitemNoStop [0]{.\EOS\space}%
\providecommand \EOS [0]{\spacefactor3000\relax}%
\providecommand \BibitemShut  [1]{\csname bibitem#1\endcsname}%
\let\auto@bib@innerbib\@empty
\bibitem [{\citenamefont {Zeldovich}\ and\ \citenamefont
  {Starobinsky}(1971)}]{Zeldovich:1971mw}%
  \BibitemOpen
  \bibfield  {author} {\bibinfo {author} {\bibfnamefont {Y.~B.}\ \bibnamefont
  {Zeldovich}}\ and\ \bibinfo {author} {\bibfnamefont {A.~A.}\ \bibnamefont
  {Starobinsky}},\ }\bibfield  {title} {\bibinfo {title} {{Particle production
  and vacuum polarization in an anisotropic gravitational field}},\ }\href@noop
  {} {\bibfield  {journal} {\bibinfo  {journal} {Zh. Eksp. Teor. Fiz.}\
  }\textbf {\bibinfo {volume} {61}},\ \bibinfo {pages} {2161} (\bibinfo {year}
  {1971})}\BibitemShut {NoStop}%
\bibitem [{\citenamefont {Unruh}(1974)}]{Unruh:1974bw}%
  \BibitemOpen
  \bibfield  {author} {\bibinfo {author} {\bibfnamefont {W.~G.}\ \bibnamefont
  {Unruh}},\ }\bibfield  {title} {\bibinfo {title} {{Second quantization in the
  Kerr metric}},\ }\href {https://doi.org/10.1103/PhysRevD.10.3194} {\bibfield
  {journal} {\bibinfo  {journal} {Phys. Rev. D}\ }\textbf {\bibinfo {volume}
  {10}},\ \bibinfo {pages} {3194} (\bibinfo {year} {1974})}\BibitemShut
  {NoStop}%
\bibitem [{\citenamefont {Hawking}(1975)}]{Hawking:1975vcx}%
  \BibitemOpen
  \bibfield  {author} {\bibinfo {author} {\bibfnamefont {S.~W.}\ \bibnamefont
  {Hawking}},\ }\bibfield  {title} {\bibinfo {title} {{Particle Creation by
  Black Holes}},\ }\href {https://doi.org/10.1007/BF02345020} {\bibfield
  {journal} {\bibinfo  {journal} {Commun. Math. Phys.}\ }\textbf {\bibinfo
  {volume} {43}},\ \bibinfo {pages} {199} (\bibinfo {year} {1975})},\ \bibinfo
  {note} {[Erratum: Commun.Math.Phys. 46, 206 (1976)]}\BibitemShut {NoStop}%
\bibitem [{\citenamefont {Zilberman}\ \emph {et~al.}(2020)\citenamefont
  {Zilberman}, \citenamefont {Levi},\ and\ \citenamefont
  {Ori}}]{Zilberman:2019buh}%
  \BibitemOpen
  \bibfield  {author} {\bibinfo {author} {\bibfnamefont {N.}~\bibnamefont
  {Zilberman}}, \bibinfo {author} {\bibfnamefont {A.}~\bibnamefont {Levi}},\
  and\ \bibinfo {author} {\bibfnamefont {A.}~\bibnamefont {Ori}},\ }\bibfield
  {title} {\bibinfo {title} {{Quantum fluxes at the inner horizon of a
  spherical charged black hole}},\ }\href
  {https://doi.org/10.1103/PhysRevLett.124.171302} {\bibfield  {journal}
  {\bibinfo  {journal} {Phys. Rev. Lett.}\ }\textbf {\bibinfo {volume} {124}},\
  \bibinfo {pages} {171302} (\bibinfo {year} {2020})},\ \Eprint
  {https://arxiv.org/abs/1906.11303} {arXiv:1906.11303 [gr-qc]} \BibitemShut
  {NoStop}%
\bibitem [{\citenamefont {Hollands}\ \emph {et~al.}(2020)\citenamefont
  {Hollands}, \citenamefont {Wald},\ and\ \citenamefont
  {Zahn}}]{Hollands:2019whz}%
  \BibitemOpen
  \bibfield  {author} {\bibinfo {author} {\bibfnamefont {S.}~\bibnamefont
  {Hollands}}, \bibinfo {author} {\bibfnamefont {R.~M.}\ \bibnamefont {Wald}},\
  and\ \bibinfo {author} {\bibfnamefont {J.}~\bibnamefont {Zahn}},\ }\bibfield
  {title} {\bibinfo {title} {{Quantum instability of the Cauchy horizon in
  Reissner\textendash{}Nordstr\"om\textendash{}deSitter spacetime}},\ }\href
  {https://doi.org/10.1088/1361-6382/ab8052} {\bibfield  {journal} {\bibinfo
  {journal} {Class. Quant. Grav.}\ }\textbf {\bibinfo {volume} {37}},\ \bibinfo
  {pages} {115009} (\bibinfo {year} {2020})},\ \Eprint
  {https://arxiv.org/abs/1912.06047} {arXiv:1912.06047 [gr-qc]} \BibitemShut
  {NoStop}%
\bibitem [{\citenamefont {Klein}\ \emph {et~al.}(2021)\citenamefont {Klein},
  \citenamefont {Zahn},\ and\ \citenamefont {Hollands}}]{Klein:2021ctt}%
  \BibitemOpen
  \bibfield  {author} {\bibinfo {author} {\bibfnamefont {C.}~\bibnamefont
  {Klein}}, \bibinfo {author} {\bibfnamefont {J.}~\bibnamefont {Zahn}},\ and\
  \bibinfo {author} {\bibfnamefont {S.}~\bibnamefont {Hollands}},\ }\bibfield
  {title} {\bibinfo {title} {{Quantum (Dis)Charge of Black Hole Interiors}},\
  }\href {https://doi.org/10.1103/PhysRevLett.127.231301} {\bibfield  {journal}
  {\bibinfo  {journal} {Phys. Rev. Lett.}\ }\textbf {\bibinfo {volume} {127}},\
  \bibinfo {pages} {231301} (\bibinfo {year} {2021})},\ \Eprint
  {https://arxiv.org/abs/2103.03714} {arXiv:2103.03714 [gr-qc]} \BibitemShut
  {NoStop}%
\bibitem [{\citenamefont {Zilberman}\ \emph {et~al.}(2022)\citenamefont
  {Zilberman}, \citenamefont {Casals}, \citenamefont {Ori},\ and\ \citenamefont
  {Ottewill}}]{Zilberman2022}%
  \BibitemOpen
  \bibfield  {author} {\bibinfo {author} {\bibfnamefont {N.}~\bibnamefont
  {Zilberman}}, \bibinfo {author} {\bibfnamefont {M.}~\bibnamefont {Casals}},
  \bibinfo {author} {\bibfnamefont {A.}~\bibnamefont {Ori}},\ and\ \bibinfo
  {author} {\bibfnamefont {A.~C.}\ \bibnamefont {Ottewill}},\ }\bibfield
  {title} {\bibinfo {title} {{Quantum Fluxes at the Inner Horizon of a Spinning
  Black Hole}},\ }\href {https://doi.org/10.1103/PhysRevLett.129.261102}
  {\bibfield  {journal} {\bibinfo  {journal} {Phys. Rev. Lett.}\ }\textbf
  {\bibinfo {volume} {129}},\ \bibinfo {pages} {261102} (\bibinfo {year}
  {2022})}\BibitemShut {NoStop}%
\bibitem [{\citenamefont {Penrose}(1974)}]{Penrose}%
  \BibitemOpen
  \bibfield  {author} {\bibinfo {author} {\bibfnamefont {R.}~\bibnamefont
  {Penrose}},\ }\bibfield  {title} {\bibinfo {title} {Gravitational collapse},\
  }in\ \href@noop {} {\emph {\bibinfo {booktitle} {Gravitational Radiation and
  Gravitational Collapse}}},\ \bibinfo {editor} {edited by\ \bibinfo {editor}
  {\bibfnamefont {C.}~\bibnamefont {DeWitt-Morette}}}\ (\bibinfo  {publisher}
  {Springer},\ \bibinfo {year} {1974})\BibitemShut {NoStop}%
\bibitem [{\citenamefont {Sorce}\ and\ \citenamefont {Wald}(2017)}]{Sorce2017}%
  \BibitemOpen
  \bibfield  {author} {\bibinfo {author} {\bibfnamefont {J.}~\bibnamefont
  {Sorce}}\ and\ \bibinfo {author} {\bibfnamefont {R.~M.}\ \bibnamefont
  {Wald}},\ }\bibfield  {title} {\bibinfo {title} {{Gedanken experiments to
  destroy a black hole. II. Kerr-Newman black holes cannot be overcharged or
  overspun}},\ }\href {https://doi.org/10.1103/PhysRevD.96.104014} {\bibfield
  {journal} {\bibinfo  {journal} {Phys. Rev. D}\ }\textbf {\bibinfo {volume}
  {96}},\ \bibinfo {pages} {104014} (\bibinfo {year} {2017})}\BibitemShut
  {NoStop}%
\bibitem [{\citenamefont {Candelas}\ \emph {et~al.}(1981)\citenamefont
  {Candelas}, \citenamefont {Chrzanowski},\ and\ \citenamefont
  {Howard}}]{Candelas1981}%
  \BibitemOpen
  \bibfield  {author} {\bibinfo {author} {\bibfnamefont {P.}~\bibnamefont
  {Candelas}}, \bibinfo {author} {\bibfnamefont {P.}~\bibnamefont
  {Chrzanowski}},\ and\ \bibinfo {author} {\bibfnamefont {K.~W.}\ \bibnamefont
  {Howard}},\ }\bibfield  {title} {\bibinfo {title} {{Quantization of
  electromagnetic and gravitational perturbations of a Kerr black hole}},\
  }\href {https://doi.org/10.1103/PhysRevD.24.297} {\bibfield  {journal}
  {\bibinfo  {journal} {Phys. Rev. D}\ }\textbf {\bibinfo {volume} {24}},\
  \bibinfo {pages} {297} (\bibinfo {year} {1981})}\BibitemShut {NoStop}%
\bibitem [{\citenamefont {Teukolsky}(1973)}]{Teukolsky1973}%
  \BibitemOpen
  \bibfield  {author} {\bibinfo {author} {\bibfnamefont {S.~A.}\ \bibnamefont
  {Teukolsky}},\ }\bibfield  {title} {\bibinfo {title} {{Perturbations of a
  Rotating Black Hole. I. Fundamental Equations for Gravitational,
  Electromagnetic, and Neutrino-Field Perturbations}},\ }\href
  {https://doi.org/10.1086/152444} {\bibfield  {journal} {\bibinfo  {journal}
  {Astrophys. J.}\ }\textbf {\bibinfo {volume} {185}},\ \bibinfo {pages} {635}
  (\bibinfo {year} {1973})}\BibitemShut {NoStop}%
\bibitem [{\citenamefont {Jensen}\ \emph {et~al.}(1995)\citenamefont {Jensen},
  \citenamefont {{Mc Laughlin}},\ and\ \citenamefont {Ottewill}}]{Jensen1995}%
  \BibitemOpen
  \bibfield  {author} {\bibinfo {author} {\bibfnamefont {B.~P.}\ \bibnamefont
  {Jensen}}, \bibinfo {author} {\bibfnamefont {J.~G.}\ \bibnamefont {{Mc
  Laughlin}}},\ and\ \bibinfo {author} {\bibfnamefont {A.~C.}\ \bibnamefont
  {Ottewill}},\ }\bibfield  {title} {\bibinfo {title} {{One-loop quantum
  gravity in Schwarzschild space-time}},\ }\href
  {https://doi.org/10.1103/PhysRevD.51.5676} {\bibfield  {journal} {\bibinfo
  {journal} {Phys. Rev. D}\ }\textbf {\bibinfo {volume} {51}},\ \bibinfo
  {pages} {5676} (\bibinfo {year} {1995})}\BibitemShut {NoStop}%
\bibitem [{\citenamefont {Casals}\ and\ \citenamefont
  {Ottewill}(2005)}]{Casals2005}%
  \BibitemOpen
  \bibfield  {author} {\bibinfo {author} {\bibfnamefont {M.}~\bibnamefont
  {Casals}}\ and\ \bibinfo {author} {\bibfnamefont {A.~C.}\ \bibnamefont
  {Ottewill}},\ }\bibfield  {title} {\bibinfo {title} {{Canonical quantization
  of the electromagnetic field on the Kerr background}},\ }\href
  {https://doi.org/10.1103/PhysRevD.71.124016} {\bibfield  {journal} {\bibinfo
  {journal} {Phys. Rev. D}\ }\textbf {\bibinfo {volume} {71}},\ \bibinfo
  {pages} {124016} (\bibinfo {year} {2005})}\BibitemShut {NoStop}%
\bibitem [{\citenamefont {Wald}(1994)}]{WaldQFTCS}%
  \BibitemOpen
  \bibfield  {author} {\bibinfo {author} {\bibfnamefont {R.~M.}\ \bibnamefont
  {Wald}},\ }\href@noop {} {\emph {\bibinfo {title} {{Quantum Field Theory in
  Curved Spacetime and Black Hole Thermodynamics}}}}\ (\bibinfo  {publisher}
  {The University of Chicago Press},\ \bibinfo {year} {1994})\BibitemShut
  {NoStop}%
\bibitem [{\citenamefont {Hollands}\ and\ \citenamefont
  {Wald}(2015)}]{Hollands2015}%
  \BibitemOpen
  \bibfield  {author} {\bibinfo {author} {\bibfnamefont {S.}~\bibnamefont
  {Hollands}}\ and\ \bibinfo {author} {\bibfnamefont {R.~M.}\ \bibnamefont
  {Wald}},\ }\bibfield  {title} {\bibinfo {title} {{Quantum fields in curved
  spacetime}},\ }\href {https://doi.org/10.1016/j.physrep.2015.02.001}
  {\bibfield  {journal} {\bibinfo  {journal} {Phys. Rep.}\ }\textbf {\bibinfo
  {volume} {574}},\ \bibinfo {pages} {1} (\bibinfo {year} {2015})},\ \Eprint
  {https://arxiv.org/abs/1401.2026} {arXiv:1401.2026} \BibitemShut {NoStop}%
\bibitem [{\citenamefont {T{\'{o}}th}(2018)}]{Toth2018}%
  \BibitemOpen
  \bibfield  {author} {\bibinfo {author} {\bibfnamefont {G.~Z.}\ \bibnamefont
  {T{\'{o}}th}},\ }\bibfield  {title} {\bibinfo {title} {{Noether currents for
  the Teukolsky master equation}},\ }\href
  {https://doi.org/10.1088/1361-6382/aad712} {\bibfield  {journal} {\bibinfo
  {journal} {Class. Quantum Gravity}\ }\textbf {\bibinfo {volume} {35}},\
  \bibinfo {pages} {185009} (\bibinfo {year} {2018})}\BibitemShut {NoStop}%
\bibitem [{\citenamefont {Balakumar}\ and\ \citenamefont
  {Winstanley}(2020)}]{Balakumar2020}%
  \BibitemOpen
  \bibfield  {author} {\bibinfo {author} {\bibfnamefont {V.}~\bibnamefont
  {Balakumar}}\ and\ \bibinfo {author} {\bibfnamefont {E.}~\bibnamefont
  {Winstanley}},\ }\bibfield  {title} {\bibinfo {title} {{Hadamard
  renormalization for a charged scalar field}},\ }\href
  {https://doi.org/10.1088/1361-6382/ab6b6e} {\bibfield  {journal} {\bibinfo
  {journal} {Class. Quantum Gravity}\ }\textbf {\bibinfo {volume} {37}},\
  \bibinfo {pages} {065004} (\bibinfo {year} {2020})}\BibitemShut {NoStop}%
\bibitem [{\citenamefont {Geroch}\ \emph {et~al.}(1973)\citenamefont {Geroch},
  \citenamefont {Held},\ and\ \citenamefont {Penrose}}]{Geroch1973}%
  \BibitemOpen
  \bibfield  {author} {\bibinfo {author} {\bibfnamefont {R.}~\bibnamefont
  {Geroch}}, \bibinfo {author} {\bibfnamefont {A.}~\bibnamefont {Held}},\ and\
  \bibinfo {author} {\bibfnamefont {R.}~\bibnamefont {Penrose}},\ }\bibfield
  {title} {\bibinfo {title} {{A space-time calculus based on pairs of null
  directions}},\ }\href {https://doi.org/10.1063/1.1666410} {\bibfield
  {journal} {\bibinfo  {journal} {J. Math. Phys.}\ }\textbf {\bibinfo {volume}
  {14}},\ \bibinfo {pages} {874} (\bibinfo {year} {1973})}\BibitemShut
  {NoStop}%
\bibitem [{\citenamefont {Pound}\ and\ \citenamefont
  {Wardell}(2021)}]{Pound2021}%
  \BibitemOpen
  \bibfield  {author} {\bibinfo {author} {\bibfnamefont {A.}~\bibnamefont
  {Pound}}\ and\ \bibinfo {author} {\bibfnamefont {B.}~\bibnamefont
  {Wardell}},\ }\bibfield  {title} {\bibinfo {title} {{Black Hole Perturbation
  Theory and Gravitational Self-Force}},\ }in\ \href
  {https://doi.org/10.1007/978-981-15-4702-7_38-1} {\emph {\bibinfo {booktitle}
  {Handb. Gravitational Wave Astron.}}}\ (\bibinfo  {publisher} {Springer
  Singapore},\ \bibinfo {address} {Singapore},\ \bibinfo {year} {2021})\ pp.\
  \bibinfo {pages} {1--119},\ \Eprint {https://arxiv.org/abs/2101.04592}
  {arXiv:2101.04592} \BibitemShut {NoStop}%
\bibitem [{\citenamefont {Wald}(1978)}]{Wald78}%
  \BibitemOpen
  \bibfield  {author} {\bibinfo {author} {\bibfnamefont {R.~M.}\ \bibnamefont
  {Wald}},\ }\bibfield  {title} {\bibinfo {title} {{Construction of Solutions
  of Gravitational, Electromagnetic, or Other Perturbation Equations from
  Solutions of Decoupled Equations}},\ }\href
  {https://doi.org/10.1103/PhysRevLett.41.203} {\bibfield  {journal} {\bibinfo
  {journal} {Phys. Rev. Lett.}\ }\textbf {\bibinfo {volume} {41}},\ \bibinfo
  {pages} {203} (\bibinfo {year} {1978})}\BibitemShut {NoStop}%
\bibitem [{\citenamefont {Chrzanowski}(1975)}]{Chrzanowski1975}%
  \BibitemOpen
  \bibfield  {author} {\bibinfo {author} {\bibfnamefont {P.~L.}\ \bibnamefont
  {Chrzanowski}},\ }\bibfield  {title} {\bibinfo {title} {{Vector potential and
  metric perturbations of a rotating black hole}},\ }\href
  {https://doi.org/10.1103/PhysRevD.11.2042} {\bibfield  {journal} {\bibinfo
  {journal} {Phys. Rev. D}\ }\textbf {\bibinfo {volume} {11}},\ \bibinfo
  {pages} {2042} (\bibinfo {year} {1975})}\BibitemShut {NoStop}%
\bibitem [{\citenamefont {Cohen}\ and\ \citenamefont
  {Kegeles}(1974)}]{Cohen1974}%
  \BibitemOpen
  \bibfield  {author} {\bibinfo {author} {\bibfnamefont {J.~M.}\ \bibnamefont
  {Cohen}}\ and\ \bibinfo {author} {\bibfnamefont {L.~S.}\ \bibnamefont
  {Kegeles}},\ }\bibfield  {title} {\bibinfo {title} {{Electromagnetic fields
  in curved spaces: A constructive procedure}},\ }\href
  {https://doi.org/10.1103/PhysRevD.10.1070} {\bibfield  {journal} {\bibinfo
  {journal} {Phys. Rev. D}\ }\textbf {\bibinfo {volume} {10}},\ \bibinfo
  {pages} {1070} (\bibinfo {year} {1974})}\BibitemShut {NoStop}%
\bibitem [{\citenamefont {Kinnersley}(1969)}]{Kinnersley}%
  \BibitemOpen
  \bibfield  {author} {\bibinfo {author} {\bibfnamefont {W.}~\bibnamefont
  {Kinnersley}},\ }\bibfield  {title} {\bibinfo {title} {{Type D vacuum
  metrics}},\ }\href {https://doi.org/10.1063/1.1664958} {\bibfield  {journal}
  {\bibinfo  {journal} {J. Math. Phys.}\ }\textbf {\bibinfo {volume} {10}},\
  \bibinfo {pages} {1195} (\bibinfo {year} {1969})}\BibitemShut {NoStop}%
\bibitem [{\citenamefont {Ottewill}\ and\ \citenamefont
  {Winstanley}(2000)}]{Ottewill2000}%
  \BibitemOpen
  \bibfield  {author} {\bibinfo {author} {\bibfnamefont {A.~C.}\ \bibnamefont
  {Ottewill}}\ and\ \bibinfo {author} {\bibfnamefont {E.}~\bibnamefont
  {Winstanley}},\ }\bibfield  {title} {\bibinfo {title} {{Renormalized stress
  tensor in Kerr space-time: General results}},\ }\href
  {https://doi.org/10.1103/PhysRevD.62.084018} {\bibfield  {journal} {\bibinfo
  {journal} {Phys. Rev. D}\ }\textbf {\bibinfo {volume} {62}},\ \bibinfo
  {pages} {1} (\bibinfo {year} {2000})},\ \Eprint
  {https://arxiv.org/abs/0004022} {arXiv:0004022 [gr-qc]} \BibitemShut
  {NoStop}%
\bibitem [{\citenamefont {Chandrasekhar}(1978)}]{Chandrasekhar1978}%
  \BibitemOpen
  \bibfield  {author} {\bibinfo {author} {\bibfnamefont {S.}~\bibnamefont
  {Chandrasekhar}},\ }\bibfield  {title} {\bibinfo {title} {{The gravitational
  perturbations of the Kerr black hole I. The perturbations in the quantities
  which vanish in the stationary state}},\ }\href
  {https://doi.org/10.1098/rspa.1978.0020} {\bibfield  {journal} {\bibinfo
  {journal} {Proc. R. Soc. London. A. Math. Phys. Sci.}\ }\textbf {\bibinfo
  {volume} {358}},\ \bibinfo {pages} {421} (\bibinfo {year}
  {1978})}\BibitemShut {NoStop}%
\bibitem [{\citenamefont {Casals}\ and\ \citenamefont {{Teixeira da
  Costa}}(2021)}]{Casals2021}%
  \BibitemOpen
  \bibfield  {author} {\bibinfo {author} {\bibfnamefont {M.}~\bibnamefont
  {Casals}}\ and\ \bibinfo {author} {\bibfnamefont {R.}~\bibnamefont {{Teixeira
  da Costa}}},\ }\bibfield  {title} {\bibinfo {title} {{The
  Teukolsky–Starobinsky constants: facts and fictions}},\ }\href
  {https://doi.org/10.1088/1361-6382/ac11a8} {\bibfield  {journal} {\bibinfo
  {journal} {Class. Quantum Gravity}\ }\textbf {\bibinfo {volume} {38}},\
  \bibinfo {pages} {165016} (\bibinfo {year} {2021})}\BibitemShut {NoStop}%
\bibitem [{\citenamefont {Klein}(2023)}]{Klein2023}%
  \BibitemOpen
  \bibfield  {author} {\bibinfo {author} {\bibfnamefont {C.~K.~M.}\
  \bibnamefont {Klein}},\ }\bibfield  {title} {\bibinfo {title} {{Construction
  of the Unruh State for a Real Scalar Field on the Kerr-de Sitter
  Spacetime}},\ }\bibfield  {journal} {\bibinfo  {journal} {Ann. Henri
  Poincar{\'{e}}}\ }\href {https://doi.org/10.1007/s00023-023-01273-6}
  {10.1007/s00023-023-01273-6} (\bibinfo {year} {2023})\BibitemShut {NoStop}%
\bibitem [{\citenamefont {G{\'{e}}rard}\ \emph {et~al.}(2020)\citenamefont
  {G{\'{e}}rard}, \citenamefont {H{\"{a}}fner},\ and\ \citenamefont
  {Wrochna}}]{Gerard2020}%
  \BibitemOpen
  \bibfield  {author} {\bibinfo {author} {\bibfnamefont {C.}~\bibnamefont
  {G{\'{e}}rard}}, \bibinfo {author} {\bibfnamefont {D.}~\bibnamefont
  {H{\"{a}}fner}},\ and\ \bibinfo {author} {\bibfnamefont {M.}~\bibnamefont
  {Wrochna}},\ }\bibfield  {title} {\bibinfo {title} {{The Unruh state for
  massless fermions on Kerr spacetime and its Hadamard property}},\ }\href
  {http://arxiv.org/abs/2008.10995} {\  (\bibinfo {year} {2020})},\ \Eprint
  {https://arxiv.org/abs/2008.10995} {arXiv:2008.10995} \BibitemShut {NoStop}%
\bibitem [{\citenamefont {Ori}(2003)}]{Ori2003}%
  \BibitemOpen
  \bibfield  {author} {\bibinfo {author} {\bibfnamefont {A.}~\bibnamefont
  {Ori}},\ }\bibfield  {title} {\bibinfo {title} {{Reconstruction of
  inhomogeneous metric perturbations and electromagnetic four-potential in Kerr
  spacetime}},\ }\href {https://doi.org/10.1103/PhysRevD.67.124010} {\bibfield
  {journal} {\bibinfo  {journal} {Phys. Rev. D}\ }\textbf {\bibinfo {volume}
  {67}},\ \bibinfo {pages} {124010} (\bibinfo {year} {2003})},\ \Eprint
  {https://arxiv.org/abs/0207045} {arXiv:0207045 [gr-qc]} \BibitemShut
  {NoStop}%
\bibitem [{\citenamefont {Frolov}\ and\ \citenamefont
  {Thorne}(1989)}]{Frolov1989}%
  \BibitemOpen
  \bibfield  {author} {\bibinfo {author} {\bibfnamefont {V.~P.}\ \bibnamefont
  {Frolov}}\ and\ \bibinfo {author} {\bibfnamefont {K.~S.}\ \bibnamefont
  {Thorne}},\ }\bibfield  {title} {\bibinfo {title} {{Renormalized
  stress-energy tensor near the horizon of a slowly evolving, rotating black
  hole}},\ }\href {https://doi.org/10.1103/PhysRevD.39.2125} {\bibfield
  {journal} {\bibinfo  {journal} {Phys. Rev. D}\ }\textbf {\bibinfo {volume}
  {39}},\ \bibinfo {pages} {2125} (\bibinfo {year} {1989})}\BibitemShut
  {NoStop}%
\bibitem [{\citenamefont {Hawking}\ and\ \citenamefont
  {Hartle}(1972)}]{Hawking1972}%
  \BibitemOpen
  \bibfield  {author} {\bibinfo {author} {\bibfnamefont {S.~W.}\ \bibnamefont
  {Hawking}}\ and\ \bibinfo {author} {\bibfnamefont {J.~B.}\ \bibnamefont
  {Hartle}},\ }\bibfield  {title} {\bibinfo {title} {{Energy and angular
  momentum flow into a black hole}},\ }\href
  {https://doi.org/10.1007/BF01645515} {\bibfield  {journal} {\bibinfo
  {journal} {Commun. Math. Phys.}\ }\textbf {\bibinfo {volume} {27}},\ \bibinfo
  {pages} {283} (\bibinfo {year} {1972})}\BibitemShut {NoStop}%
\bibitem [{\citenamefont {Teukolsky}\ and\ \citenamefont
  {Press}(1974)}]{Teukolsky1974}%
  \BibitemOpen
  \bibfield  {author} {\bibinfo {author} {\bibfnamefont {S.~A.}\ \bibnamefont
  {Teukolsky}}\ and\ \bibinfo {author} {\bibfnamefont {W.~H.}\ \bibnamefont
  {Press}},\ }\bibfield  {title} {\bibinfo {title} {{Perturbations of a
  rotating black hole. III - Interaction of the hole with gravitational and
  electromagnetic radiation}},\ }\href {https://doi.org/10.1086/153180}
  {\bibfield  {journal} {\bibinfo  {journal} {Astrophys. J.}\ }\textbf
  {\bibinfo {volume} {193}},\ \bibinfo {pages} {443} (\bibinfo {year}
  {1974})}\BibitemShut {NoStop}%
\bibitem [{\citenamefont {Klein}\ and\ \citenamefont {Zahn}(2021)}]{Klein2021}%
  \BibitemOpen
  \bibfield  {author} {\bibinfo {author} {\bibfnamefont {C.}~\bibnamefont
  {Klein}}\ and\ \bibinfo {author} {\bibfnamefont {J.}~\bibnamefont {Zahn}},\
  }\bibfield  {title} {\bibinfo {title} {{Renormalized charged scalar current
  in the Reissner–Nordstr{\"{o}}m–de Sitter spacetime}},\ }\href
  {https://doi.org/10.1103/PhysRevD.104.025009} {\bibfield  {journal} {\bibinfo
   {journal} {Phys. Rev. D}\ }\textbf {\bibinfo {volume} {104}},\ \bibinfo
  {pages} {025009} (\bibinfo {year} {2021})}\BibitemShut {NoStop}%
\bibitem [{\citenamefont {Wald}(1973)}]{Wald1973}%
  \BibitemOpen
  \bibfield  {author} {\bibinfo {author} {\bibfnamefont {R.~M.}\ \bibnamefont
  {Wald}},\ }\bibfield  {title} {\bibinfo {title} {{On perturbations of a Kerr
  black hole}},\ }\href {https://doi.org/10.1063/1.1666203} {\bibfield
  {journal} {\bibinfo  {journal} {J. Math. Phys.}\ }\textbf {\bibinfo {volume}
  {14}},\ \bibinfo {pages} {1453} (\bibinfo {year} {1973})}\BibitemShut
  {NoStop}%
\bibitem [{\citenamefont {Green}\ \emph {et~al.}(2020)\citenamefont {Green},
  \citenamefont {Hollands},\ and\ \citenamefont {Zimmerman}}]{Green2020}%
  \BibitemOpen
  \bibfield  {author} {\bibinfo {author} {\bibfnamefont {S.~R.}\ \bibnamefont
  {Green}}, \bibinfo {author} {\bibfnamefont {S.}~\bibnamefont {Hollands}},\
  and\ \bibinfo {author} {\bibfnamefont {P.}~\bibnamefont {Zimmerman}},\
  }\bibfield  {title} {\bibinfo {title} {{Teukolsky formalism for nonlinear
  Kerr perturbations}},\ }\href {https://doi.org/10.1088/1361-6382/ab7075}
  {\bibfield  {journal} {\bibinfo  {journal} {Class. Quantum Gravity}\ }\textbf
  {\bibinfo {volume} {37}},\ \bibinfo {pages} {1} (\bibinfo {year} {2020})},\
  \Eprint {https://arxiv.org/abs/1908.09095} {arXiv:1908.09095} \BibitemShut
  {NoStop}%
\bibitem [{\citenamefont {Prabhu}\ and\ \citenamefont
  {Wald}(2018)}]{Prabhu2018}%
  \BibitemOpen
  \bibfield  {author} {\bibinfo {author} {\bibfnamefont {K.}~\bibnamefont
  {Prabhu}}\ and\ \bibinfo {author} {\bibfnamefont {R.~M.}\ \bibnamefont
  {Wald}},\ }\bibfield  {title} {\bibinfo {title} {{Canonical energy and Hertz
  potentials for perturbations of Schwarzschild spacetime}},\ }\href
  {https://doi.org/10.1088/1361-6382/aae9ae} {\bibfield  {journal} {\bibinfo
  {journal} {Class. Quantum Gravity}\ }\textbf {\bibinfo {volume} {35}},\
  \bibinfo {pages} {1} (\bibinfo {year} {2018})},\ \Eprint
  {https://arxiv.org/abs/1807.09883} {arXiv:1807.09883} \BibitemShut {NoStop}%
\bibitem [{\citenamefont {Poisson}\ \emph {et~al.}(2011)\citenamefont
  {Poisson}, \citenamefont {Pound},\ and\ \citenamefont {Vega}}]{Poisson2011}%
  \BibitemOpen
  \bibfield  {author} {\bibinfo {author} {\bibfnamefont {E.}~\bibnamefont
  {Poisson}}, \bibinfo {author} {\bibfnamefont {A.}~\bibnamefont {Pound}},\
  and\ \bibinfo {author} {\bibfnamefont {I.}~\bibnamefont {Vega}},\ }\bibfield
  {title} {\bibinfo {title} {{The motion of point particles in curved
  spacetime}},\ }\href {https://doi.org/10.12942/lrr-2011-7} {\bibfield
  {journal} {\bibinfo  {journal} {Living Rev. Relativ.}\ }\textbf {\bibinfo
  {volume} {14}},\ \bibinfo {pages} {1} (\bibinfo {year} {2011})},\ \Eprint
  {https://arxiv.org/abs/1102.0529} {arXiv:1102.0529} \BibitemShut {NoStop}%
\bibitem [{\citenamefont {Aksteiner}\ and\ \citenamefont
  {B{\"{a}}ckdahl}(2018)}]{Aksteiner2018}%
  \BibitemOpen
  \bibfield  {author} {\bibinfo {author} {\bibfnamefont {S.}~\bibnamefont
  {Aksteiner}}\ and\ \bibinfo {author} {\bibfnamefont {T.}~\bibnamefont
  {B{\"{a}}ckdahl}},\ }\bibfield  {title} {\bibinfo {title} {{All Local Gauge
  Invariants for Perturbations of the Kerr Spacetime}},\ }\href
  {https://doi.org/10.1103/PhysRevLett.121.051104} {\bibfield  {journal}
  {\bibinfo  {journal} {Phys. Rev. Lett.}\ }\textbf {\bibinfo {volume} {121}},\
  \bibinfo {pages} {051104} (\bibinfo {year} {2018})}\BibitemShut {NoStop}%
\bibitem [{\citenamefont {Hollands}\ and\ \citenamefont
  {Wald}(2001)}]{Hollands2001}%
  \BibitemOpen
  \bibfield  {author} {\bibinfo {author} {\bibfnamefont {S.}~\bibnamefont
  {Hollands}}\ and\ \bibinfo {author} {\bibfnamefont {R.~M.}\ \bibnamefont
  {Wald}},\ }\bibfield  {title} {\bibinfo {title} {{Local Wick Polynomials and
  Time Ordered Products of Quantum Fields in Curved Spacetime}},\ }\href
  {https://doi.org/10.1007/s002200100540} {\bibfield  {journal} {\bibinfo
  {journal} {Commun. Math. Phys.}\ }\textbf {\bibinfo {volume} {223}},\
  \bibinfo {pages} {289} (\bibinfo {year} {2001})}\BibitemShut {NoStop}%
\bibitem [{\citenamefont {Ottewill}\ and\ \citenamefont
  {Wardell}(2009)}]{Ottewill2009}%
  \BibitemOpen
  \bibfield  {author} {\bibinfo {author} {\bibfnamefont {A.~C.}\ \bibnamefont
  {Ottewill}}\ and\ \bibinfo {author} {\bibfnamefont {B.}~\bibnamefont
  {Wardell}},\ }\bibfield  {title} {\bibinfo {title} {{Quasilocal contribution
  to the scalar self-force: Nongeodesic motion}},\ }\href
  {https://doi.org/10.1103/PhysRevD.79.024031} {\bibfield  {journal} {\bibinfo
  {journal} {Phys. Rev. D}\ }\textbf {\bibinfo {volume} {79}},\ \bibinfo
  {pages} {024031} (\bibinfo {year} {2009})},\ \Eprint
  {https://arxiv.org/abs/0810.1961} {arXiv:0810.1961} \BibitemShut {NoStop}%
\bibitem [{\citenamefont {Levi}\ and\ \citenamefont {Ori}(2015)}]{Levi2015}%
  \BibitemOpen
  \bibfield  {author} {\bibinfo {author} {\bibfnamefont {A.}~\bibnamefont
  {Levi}}\ and\ \bibinfo {author} {\bibfnamefont {A.}~\bibnamefont {Ori}},\
  }\bibfield  {title} {\bibinfo {title} {{Pragmatic mode-sum regularization
  method for semiclassical black-hole spacetimes}},\ }\href
  {https://doi.org/10.1103/PhysRevD.91.104028} {\bibfield  {journal} {\bibinfo
  {journal} {Phys. Rev. D}\ }\textbf {\bibinfo {volume} {91}},\ \bibinfo
  {pages} {104028} (\bibinfo {year} {2015})}\BibitemShut {NoStop}%
\bibitem [{\citenamefont {Hollands}\ and\ \citenamefont
  {Wald}(2013)}]{Hollands2013}%
  \BibitemOpen
  \bibfield  {author} {\bibinfo {author} {\bibfnamefont {S.}~\bibnamefont
  {Hollands}}\ and\ \bibinfo {author} {\bibfnamefont {R.~M.}\ \bibnamefont
  {Wald}},\ }\bibfield  {title} {\bibinfo {title} {{Stability of Black Holes
  and Black Branes}},\ }\href {https://doi.org/10.1007/s00220-012-1638-1}
  {\bibfield  {journal} {\bibinfo  {journal} {Commun. Math. Phys.}\ }\textbf
  {\bibinfo {volume} {321}},\ \bibinfo {pages} {629} (\bibinfo {year}
  {2013})},\ \Eprint {https://arxiv.org/abs/1201.0463} {arXiv:1201.0463}
  \BibitemShut {NoStop}%
\bibitem [{\citenamefont {Collinson}(1990)}]{Collinson1990}%
  \BibitemOpen
  \bibfield  {author} {\bibinfo {author} {\bibfnamefont {C.~D.}\ \bibnamefont
  {Collinson}},\ }\bibfield  {title} {\bibinfo {title} {{Tetrad symmetries}},\
  }\href {https://doi.org/10.1007/BF00759017} {\bibfield  {journal} {\bibinfo
  {journal} {Gen. Relativ. Gravit.}\ }\textbf {\bibinfo {volume} {22}},\
  \bibinfo {pages} {1163} (\bibinfo {year} {1990})}\BibitemShut {NoStop}%
\bibitem [{\citenamefont {Wernersson}\ and\ \citenamefont
  {Zahn}(2021)}]{Wernersson2021}%
  \BibitemOpen
  \bibfield  {author} {\bibinfo {author} {\bibfnamefont {J.}~\bibnamefont
  {Wernersson}}\ and\ \bibinfo {author} {\bibfnamefont {J.}~\bibnamefont
  {Zahn}},\ }\bibfield  {title} {\bibinfo {title} {{Vacuum polarization near
  boundaries}},\ }\href {https://doi.org/10.1103/PhysRevD.103.016012}
  {\bibfield  {journal} {\bibinfo  {journal} {Phys. Rev. D}\ }\textbf {\bibinfo
  {volume} {103}},\ \bibinfo {pages} {16012} (\bibinfo {year} {2021})},\
  \Eprint {https://arxiv.org/abs/2010.05499} {arXiv:2010.05499} \BibitemShut
  {NoStop}%
\bibitem [{\citenamefont {Levi}(2017)}]{Levi2017}%
  \BibitemOpen
  \bibfield  {author} {\bibinfo {author} {\bibfnamefont {A.}~\bibnamefont
  {Levi}},\ }\bibfield  {title} {\bibinfo {title} {{Renormalized stress-energy
  tensor for stationary black holes}},\ }\href
  {https://doi.org/10.1103/PhysRevD.95.025007} {\bibfield  {journal} {\bibinfo
  {journal} {Phys. Rev. D}\ }\textbf {\bibinfo {volume} {95}},\ \bibinfo
  {pages} {025007} (\bibinfo {year} {2017})}\BibitemShut {NoStop}%
\bibitem [{\citenamefont {Zahn}(2015)}]{Zahn2015}%
  \BibitemOpen
  \bibfield  {author} {\bibinfo {author} {\bibfnamefont {J.}~\bibnamefont
  {Zahn}},\ }\bibfield  {title} {\bibinfo {title} {{Locally covariant chiral
  fermions and anomalies}},\ }\href
  {https://doi.org/10.1016/j.nuclphysb.2014.11.008} {\bibfield  {journal}
  {\bibinfo  {journal} {Nucl. Phys. B}\ }\textbf {\bibinfo {volume} {890}},\
  \bibinfo {pages} {1} (\bibinfo {year} {2015})},\ \Eprint
  {https://arxiv.org/abs/1407.1994} {arXiv:1407.1994} \BibitemShut {NoStop}%
\bibitem [{\citenamefont {Balakumar}\ \emph {et~al.}(2022)\citenamefont
  {Balakumar}, \citenamefont {Bernar},\ and\ \citenamefont
  {Winstanley}}]{Balakumar2022}%
  \BibitemOpen
  \bibfield  {author} {\bibinfo {author} {\bibfnamefont {V.}~\bibnamefont
  {Balakumar}}, \bibinfo {author} {\bibfnamefont {R.~P.}\ \bibnamefont
  {Bernar}},\ and\ \bibinfo {author} {\bibfnamefont {E.}~\bibnamefont
  {Winstanley}},\ }\bibfield  {title} {\bibinfo {title} {{Quantization of a
  charged scalar field on a charged black hole background}},\ }\href
  {https://doi.org/10.1103/PhysRevD.106.125013} {\bibfield  {journal} {\bibinfo
   {journal} {Phys. Rev. D}\ }\textbf {\bibinfo {volume} {106}},\ \bibinfo
  {pages} {125013} (\bibinfo {year} {2022})}\BibitemShut {NoStop}%
\bibitem [{\citenamefont {Green}\ \emph {et~al.}(2022)\citenamefont {Green},
  \citenamefont {Hollands}, \citenamefont {Sberna}, \citenamefont {Toomani},\
  and\ \citenamefont {Zimmerman}}]{Green2022}%
  \BibitemOpen
  \bibfield  {author} {\bibinfo {author} {\bibfnamefont {S.~R.}\ \bibnamefont
  {Green}}, \bibinfo {author} {\bibfnamefont {S.}~\bibnamefont {Hollands}},
  \bibinfo {author} {\bibfnamefont {L.}~\bibnamefont {Sberna}}, \bibinfo
  {author} {\bibfnamefont {V.}~\bibnamefont {Toomani}},\ and\ \bibinfo {author}
  {\bibfnamefont {P.}~\bibnamefont {Zimmerman}},\ }\bibfield  {title} {\bibinfo
  {title} {{Conserved currents for Kerr and orthogonality of quasinormal
  modes}},\ }\href {http://arxiv.org/abs/2210.15935} {\  (\bibinfo {year}
  {2022})},\ \Eprint {https://arxiv.org/abs/2210.15935} {arXiv:2210.15935}
  \BibitemShut {NoStop}%
\bibitem [{\citenamefont {Carter}(1968)}]{Carter1968}%
  \BibitemOpen
  \bibfield  {author} {\bibinfo {author} {\bibfnamefont {B.}~\bibnamefont
  {Carter}},\ }\bibfield  {title} {\bibinfo {title} {{Global Structure of the
  Kerr Family of Gravitational Fields}},\ }\href
  {https://doi.org/10.1103/PhysRev.174.1559} {\bibfield  {journal} {\bibinfo
  {journal} {Phys. Rev.}\ }\textbf {\bibinfo {volume} {174}},\ \bibinfo {pages}
  {1559} (\bibinfo {year} {1968})}\BibitemShut {NoStop}%
\bibitem [{BHP()}]{BHPToolkit}%
  \BibitemOpen
  \href@noop {} {\bibinfo {title} {{Black Hole Perturbation Toolkit}}},\
  \bibinfo {howpublished}
  {(\href{http://bhptoolkit.org/}{bhptoolkit.org})}\BibitemShut {NoStop}%
\end{thebibliography}%
\end{document}